\documentclass[twocolumn]{aastex631}
\usepackage{graphicx}
\usepackage{hyperref}
\usepackage{mathtools}
\usepackage[mathlines]{lineno}
\usepackage{xspace}
\graphicspath{ {./Figures/} }

\usepackage{color}

\def\eg{{\it e.g., }}

\def\ie{{\it i.e., }}

\def\gsim{~\rlap{$>$}{\lower 1.0ex\hbox{$\sim$}}}
\def\lsim{~\rlap{$<$}{\lower 1.0ex\hbox{$\sim$}}}

\def\tess{{\it TESS}\xspace}
\def\kepler{{\it Kepler}\xspace}
\def\jwst{{\it JWST}\xspace}

\def\msun{{M_\odot}}
\def\rsun{{R_\odot}}
\def\lsun{{L_\odot}}

\def\vplanet{\texttt{\footnotesize{VPLanet}}\xspace}

\def\alabi{\texttt{\footnotesize{alabi}}\xspace}

\def\nbody6{\texttt{\footnotesize{NBODY6++}}\xspace}

\def\emcee{\texttt{emcee}\xspace}
\def\dynesty{\texttt{dynesty}\xspace}
\def\pymultinest{\texttt{pymultinest}\xspace}
\def\ultranest{\texttt{ultranest}\xspace}

\def\k40{{$^{40}$K}}
\def\th232{{$^{232}$Th}}

\begin{document}

\title{The Range of Cumulative XUV Flux on GJ 1132 b}

\author[0000-0001-6487-5445]{Rory Barnes}
\affiliation{Department of Astronomy, University of Washington, 3910 15th Avenue NE, Seattle, WA 98195, USA}
\email{rory@astro.washington.edu}

\author[0000-0002-8341-0376]{Laura N. R. do Amaral}
\affiliation{School of Earth and Space Exploration, Arizona State University, 781 Terrace Mall, Tempe, AZ 85287, USA}

\author[0000-0002-7961-6881]{Jessica Birky}
\affiliation{Department of Astronomy, University of Washington, 3910 15th Avenue NE, Seattle, WA 98195, USA}

\author[0000-0002-0637-835X]{James R. A. Davenport}
\affiliation{Department of Astronomy, University of Washington, 3910 15th Avenue NE, Seattle, WA 98195, USA}

\author[0000-0002-0637-835X]{Scott Engle}
\affiliation{Villanova University, Dept. of Astrophysics and Planetary Science, 800 E. Lancaster Ave., Villanova, PA 19085, USA}

\author{Megan Gialluca}
\affiliation{Department of Astronomy, University of Washington, 3910 15th Avenue NE, Seattle, WA 98195, USA}

\author[0000-0002-0637-835X]{Evgenya L. Shkolnik}
\affiliation{School of Earth and Space Exploration, Arizona State University, 781 Terrace Mall, Tempe, AZ 85287, USA}

\begin{abstract}
    We investigate the plausible history of the XUV luminosity evolution of the planet-hosting M4 star GJ 1132 ($\sim$0.2 $\msun$) to infer the cumulative incident XUV flux intercepted by the short-period ($\sim$1.6 d) Earth-sized transiting planet GJ 1132 b. We include the dominant observational uncertainties, compare two quiescent XUV luminosity evolution models, and simulate the XUV luminosity evolution from flares based on \tess data and a re-analysis of \kepler stars. We find only 4 flares in GJ 1132's \tess 123 day lightcurve, which is relatively few for M dwarfs and, in conjunction with the $\sim125$ day period, suggests that this star is many Gyr old. We find that all model permutations predict that the planet has at least a 95\% chance of receiving more than 50 times as much XUV flux as modern Earth, confirming that this planet is a good candidate for permanent atmospheric loss. We also find that an empirical XUV model for M dwarfs predicts  2-3 times more total XUV flux than a commonly used solar twin model and that the empirical model's distribution is 2--3 times narrower.  Flares contribute about 20\% of the cumulative XUV flux on planet b, which, while modest, ensures the planet lies firmly on the  atmosphere-free side of the ``cosmic shoreline.''
\end{abstract}

\section{Introduction\label{sec:intro}}

\jwst is currently searching for atmospheres of terrestrial exoplanets on short-period orbits, but it remains unclear how persistent these atmospheres may be \citep[see, \eg][]{Kreidberg19,Crossfield22,Xue2024,Mansfield24,Wachiraphan25}. In such tight quarters, the high energy radiation from the star can remove atmospheres by accelerating atoms to escape velocity, depleting the atmosphere over time. Thus, characterizing the historical incident X-ray and ultraviolet (XUV) (1-1000\AA) flux of planet-hosting stars can provide critical insight into the longevity of short-period terrestrial planet atmospheres \citep{ZahnleCatling17}. Here we examine the XUV history of the planet host GJ 1132, a $V = 12.6$ magnitude M4, $0.2 \msun$ star \citep{Bidelman85} with at least two planets, one of which, planet b, is an optimal target for \jwst observations \citep{BertaThompson15}. Our ultimate goal is to predict the total XUV energy that has struck the planet over its lifetime.

While obtaining current X-ray luminosities, which can be scaled to XUV luminosities with significant uncertainties, is relatively straight-forward \citep{MoralesCalderon2024}, the current XUV luminosity provides minimal constraints on the historical XUV luminosity. Instead, models for the XUV evolution are assumed based on observations of stars with various ages that are then used to construct a model. One commonly used method relies on observations of a handful of ``solar twins'' \citep{Ribas05}, which suggests that stellar XUV is initially a constant fraction of the total luminosity and then decays as a power law in ime \citep[see also][]{LugerBarnes15}. Alternatively, observations of stars with ages determined by association, \eg stars in open clusters or with white dwarf companions, can generate models over a wider range of stellar masses \citep{Engle24}. Empirical methods currently offer the best path to estimating a field star's XUV history and hence statistically constraining the cumulative XUV flux to impinge upon a planet.

In general the XUV luminosity may be divided into quiescent and flaring components, which do not necessarily evolve in the same way. Several studies have developed empirical models for the quiescent  \citep{Ribas05, SchneiderShkolnik18,Peacock19,RicheyYowell22,Engle24,Peacock25,France25} and flaring \citep{Loyd18,Davenport19,Ilin19,Guenther20,Feinstein20} XUV evolution over time. We will exploit these models to constrain GJ 1132's quiescent and flaring XUV evolution based on recent observation of GJ 1132's rotation period, X-Ray luminosity, and \tess flare data.

While numerous studies have computed reasonable estimates of the cumulative XUV flux that short-period planets have intercepted \citep[e.g.,][]{Pass25}, few, if any, have provided uncertainties. Given the relatively large uncertainties in the ages of most field stars, as well as the uncertainties in the parameters of XUV evolutionary models, the currently published estimates may be misleading. To begin to rectify the situation, here we introduce a framework to generate Bayesian posteriors for the cumulative XUV flux received by GJ 1132 b over its lifetime. The approach presented below can be applied to other stars, assuming the observational data exist, to provide robust constraints on any planet's cumulative XUV flux, which can help determine if \jwst observations reveal the presence of an atmosphere or not.

Table \ref{tab:starprops} lists the relevant parameters for the host star, with values taken from \citet[][hereafter Xue24]{Xue2024}, \citet[][hereafter MC24]{MoralesCalderon2024}, and \citet[][hereafter Bonfils18]{Bonfils2018}, which reveal an old M4 dwarf star that appears to be relatively quiet today. The X-ray luminosity comes from 46 ks of \textit{XMM} data (MC24). While a calm upper atmosphere bodes well for robust detections of atmospheric or surface features with \jwst, it also makes it difficult to constrain the current flaring rate. Nonetheless, as we show below, the set of observations can be combined with modern statistical methods to infer the total XUV luminosity over time and hence the cumulative XUV flux on planet b.

\begin{table}
\centering
\caption{Literature Values for Selected GJ 1132 Properties}
\begin{tabular}{lcl}
\hline\hline
Property & Value & Ref.\\
\hline
Mass & $0.1945^{+0.0048}_{-0.0046}~\msun$ & Xue24\\
Total Luminosity & $0.00477^{+0.00036}_{-0.00026}~\lsun$ & Xue24\\
Radius & $0.2211^{+0.0069}_{-0.0081}~\rsun$ & Xue24\\
Density & $25.3^{+3.1}_{-2.2}$ g/cm$^3$ & Xue24\\
Effective Temperature & $3229^{+78}_{-62}$ K & Xue24\\
X-Ray Luminosity & $(9.96\pm2.95) \times 10^{25}$ ergs/s & MC24\\
Rotation Period & $122.3^{+6.0}_{-5.0}$ days & Bonfils18\\
\hline
\end{tabular}
\label{tab:starprops}
\end{table}

In the next section, we present summaries of the physical models we used to calculate the XUV evolution of GJ 1132. In $\S$~\ref{sec:stats}, we describe how we estimate uncertainties for the different aspects of the experiment. Then we reanalyze the \kepler flare data set of 347 M dwarf stars from \citet{Davenport19}, but take into account non-detections, in $\S$~\ref{sec:kepler}.  In $\S$~\ref{sec:obsflare} we present the results of our search for flares on GJ 1132 in \tess data. $\S$~\ref{sec:quiescent} describes how we translated current X-ray luminosity into XUV luminosity. In $\S$~\ref{sec:results}, we present the results of the numerical modeling and estimate the cumulative XUV flux on GJ 1132 b for different XUV evolutionary models. We then discuss the results in $\S$~\ref{sec:discussion} and, finally, in $\S$~\ref{sec:conclusions} we summarize our findings. Appendices A and B provide additional details regarding the machine learning methods used in this investigation.

\section{Physical Models\label{sec:models}}

We use the planetary system evolution code \vplanet~\citep{Barnes20} to simulate the evolution of the star. This code includes quiescent and flaring models that predict the bolometric and XUV luminosity of the star, $L_{bol}$ and $L_{XUV}$, respectively. In the next three subsections we present the specific models we use in this study. The scripts to run the simulations and generate the figures are publicly available.\footnote{https://github.com/RoryBarnes/GJ1132\_XUV}

\subsection{Structure\label{sec:starstructure}}

We use the \cite{Baraffe15} stellar evolution models to calculate $L_{bol}$, effective temperature $T_{eff}$, and radius $R_*$ as a function of time. This model depends only on the stellar mass and metallicity. As GJ 1132's metallicity is close to solar at [Fe/H] = -0.12 \citep{BertaThompson15}, we use the solar metallicity track to model this star.

\subsection{Quiescent XUV Luminosity\label{sec:quiescencemodel}}

For the quiescent evolution, we consider two models: the ``solar twin'' model \citep[][hereafter Ribas05]{Ribas05} and the ``empirical M dwarf'' (EMD) model \citep[][]{EngleGuinan23,Engle24}. We describe these two models in the following subsubsections.

\subsubsection{The Solar Twin Model\label{sec:solartwin}}

The solar twin model was derived from 5 stars with masses and compositions that are nearly identical to the Sun, and so are significantly different from GJ 1132. Nonetheless, this model has been widely applied to M dwarfs \citep[\eg][]{LugerBarnes15,ZahnleCatling17,Mordasini20,EveOwen25,Fleming20}, so we use it here as well. In particular, we use the formulation presented in \cite{LugerBarnes15}, which consists of 5 variables: age $t$, the ``saturation time'' $t_{sat}$, the ``saturation fraction'' $f_{sat}$, the post-saturation power law exponent $\beta_{XUV}$, and the mass $m$ (which sets the bolometric luminosity over time):

\begin{align}
\label{eq:lxuv}
\frac{L_{XUV}^{quiescence}}{L_{bol}} = \left\{
        \begin{array}{lcr}
            f_\mathrm{sat} &\ & t \leq t_\mathrm{sat} \\
            f_\mathrm{sat}\left(\frac{t}{t_\mathrm{sat}}\right)^{-\beta_\mathrm{XUV}} &\ & t > t_\mathrm{sat}.
        \end{array}
    \right.
\end{align}
This model assumes that the XUV bandwidth is 1--1200\AA.

\subsubsection{Empirical M Dwarf Model\label{sec:empirical}}

The EMD model consists of a two-part linear relationship that is parameterized by the logarithm of the age in Gyr ($\equiv\tau$). The age-rotation period ($P_{rot}$) relationship is:
\begin{align}
\tau = \left\{
            \begin{array}{lcr}
                a_{rot}P_{rot} + b_{rot} &\ & P_{rot} < d_{rot} \\
                a_{rot}P_{rot} + b_{rot} + c_{rot}(P_{rot} - d_{rot}) &\ & P_{rot} \geq d_{rot},
        \end{array}
    \right.
    \label{eq:englerot}
\end{align}
when $P_{rot}$ is expressed in days. The XUV-age relationship is described by the following equation:
\begin{align}
\frac{L_{XUV}^{quiescence}}{L_{bol}} = \left\{
            \begin{array}{lcr}
                a_{xuv}\tau + b_{xuv} &\ & \tau < d_{xuv} \\
                a_{xuv}\tau + b_{xuv} + c_{xuv}(\tau - d_{xuv}) &\ & \tau \geq d_{xuv}.
        \end{array}
    \right.
    \label{eq:englexuv}
\end{align}
The coefficients are provided in Table~\ref{tab:emd} with XUV columns corresponding to the $xuv$ subscript, and the Rotation columns the $rot$ subscript. Note that for the XUV luminosity, the fits are only valid for M2 -- M6.5 stars and assume a bandpass of 5--1700\AA, whereas the rotation period fit is valid for M4 -- M6.5 stars. Below we ignore differences between the XUV bandpass widths.


\begin{deluxetable*}{c|cc|cc}
\tablecaption{Coefficients for the EMD Model \label{tab:emd}}
\tablehead{\colhead{}  & \multicolumn{2}{c}{XUV}
           & \multicolumn{2}{c}{Rotation}\\
            \hline
           \colhead{Parameter} & \colhead{Best Fit} & \colhead{Uncertainty}
           & \colhead{Best Fit} & \colhead{Uncertainty}}
\startdata
\hline
a & -0.1456 & 0.0911 & 0.0251 & 0.0018\\
b & -2.8876 & 0.0439 & -0.1615 & 0.0303\\
c  & -1.8187 & 0.2412 & -0.0212 & 0.0018\\
d & 0.3545  & 0.0604 & 25.4500 & 1.9079\\
\hline
\enddata
\end{deluxetable*}

\subsection{XUV from Flares\label{sec:flaremodel}}

Our approach to modeling the XUV output from stellar flares follows exactly the model described in \citet{Amaral22}. Stellar flares are typically described in terms of the flare frequency distribution (FFD), a power law that quantifies the frequency of flares per unit energy $\nu$:
\begin{equation}
\label{eq:ffd}
    \log(\nu) = a\log(E) + b,
\end{equation}
where $E$ is the flare energy, $a$ is the power law index (slope in log-log visualizations), and $b$ is a scaling factor (the $y$-intercept in log-log visualizations). The slope and $y$-intercept of the FFD is observed to change with time \citep[\eg][]{Davenport19}, so in general the cumulative XUV from flares must be modeled according to the time rates of change of $a$ and $b$.
The Davenport et al.~model further parameterizes the flare frequency slope and $y$-intercept as
\begin{equation}
\label{eq:ffd_a}
    a = a_1\log t + a_2 m + a_3,
\end{equation}
and
\begin{equation}
\label{eq:ffd_b}
    b = b_1\log t + b_2 m + b_3,
\end{equation}
where $m$ is the stellar mass. For more information on the details of the model and its motivation, we refer the reader to \citep{Davenport19}.

A fraction of the energy from each flare is emitted in the XUV and we use the conversion factors from
\citet[][Table 2]{OstenWolk15} to obtain this fraction. These scaling factors were obtained by assuming 9000 K blackbody flares on the Sun, which probably underestimates the flare energies. The instantaneous XUV luminosity from flares is given by
\begin{equation}
       L_{XUV}^{flares} = \int_{E_{min}}^{E_{max}} \nu(E_{XUV})\ dE,
\label{eq:LXUVFlare}
\end{equation}
where $E_{min}$ and $E_{max}$ are the minimum and maximum flare energies considered, and $E_{XUV}$ is the portion of the flare energy emitted in the XUV.

Thus, with a model for the FFD with time, we can compute the XUV luminosity from flares. The total XUV luminosity is then simply $L_{XUV}(t) = L_{XUV}^{quiescence}(t) + L_{XUV}^{flares}(t)$. For more details on this approach, consult \citet{Amaral22}.


\section{Statistical Methods\label{sec:stats}}

We employ a wide array of modern statistical tools to infer the cumulative XUV flux on GJ 1132 b, with the chosen statistical approach depending on which XUV model is used. For the EMD model, our approach is relatively simple and begins with computing the probability distribution function (PDF) of the system's age based on the rotational period constraint, Eq.~(\ref{eq:englerot}), and the values in Table~\ref{tab:emd}. The distribution of histories of the ratio of the star's quiescent XUV to bolometric luminosity is then computed via Eq.~(\ref{eq:englexuv}) and the values in Table~\ref{tab:emd}. The absolute value of the quiescent XUV luminosity over time can then be calculated using the $L_{bol}$ distribution presented in Table~\ref{tab:starprops} and \vplanet. This approach effectively constrains the plausible range of the star's quiescent XUV evolution according to the EMD model.

A similar approach, however, will not work for the Ribas05 model, which is independent of rotation. We will thus use a machine learning accelerated Markov chain Monte Carlo (MCMC) approach to infer posteriors for its 5 model parameters as conditioned on the observed bolometric and XUV luminosities; the latter is scaled from the X-ray luminosity via the \citet{SanzForcada25} scaling relationship for F to M type stars. This problem is clearly underconstrained, so prior knowledge is essential for inferring posteriors. We will use a mass prior that corresponds to the results of Xue24, the age PDF predicted by the EMD model for the age prior, and $\beta_{XUV}$, $t_{sat}$ and $f_{sat}$ priors that have been previously used for TRAPPIST-1 \citep{Fleming20,Birky21}. Note that although we use the results of the EMD model to inform the Ribas05 computation, the age-rotation relationship of \citet{EngleGuinan23} does not depend on XUV luminosity, so the two quiescent XUV models are independent.

To constrain the XUV luminosity from flares, we use an affine-invariant MCMC method (\emcee) described in the next section. We use over 485,000 data points from \kepler observations of M stars \citep{Davenport19} to infer posteriors for the $a_1, ..., b_3$ parameters of the \citet{Davenport19} model. The distributions of these parameters constrain the history of the star's XUV evolution due to flares.

With the quiescent and flaring XUV histories constrained, we may then compute their sum to derive the total XUV luminosity of the star as a function of time. We then calculate the PDF of the cumulative XUV flux planet b has received by drawing parameters from the posteriors and self-consistently computing the planet's semi-major axis from the best fit and uncertainties for the stellar mass and orbital period ($1.628931 \pm 0.000027$ days). In other words, we maintain the covariances between the stellar model parameters, however, we do not use any covariance between stellar mass and orbital period, which we expect to be independent.

We then draw from the posteriors to simulate the incident XUV flux on planet b with \vplanet over time and compute the cumulative flux by first approximating the integral between consecutive timesteps via the trapezoidal rule and then adding the result to the previous value of the cumulative flux. The end result is a Bayesian distribution of the cumulative XUV flux on planet b for various models, which can then be compared to the cosmic shoreline to calculate the probability that planet b has received sufficient XUV flux to remove its atmosphere. The following subsections provide more details on the statistical methods we use.

\subsection{Maximum Likelihood Estimation}

A crucial first step toward inferring posteriors of the Ribas05 model parameters is to calculate the  maximum \emph{a
posteriori} (MAP) estimate of the model parameters, which can serve as both an
initial reference point for the MCMC samplers and optimizing hyperparameters in our machine learning methodology. The MAP estimate is
computed with \texttt{MaxLEV}\footnote{Publicly available at https://github.com/RoryBarnes/MaxLEV}, a new maximum likelihood estimator for
\texttt{VPLanet} models. For each proposed parameter vector, \texttt{MaxLEV}
evaluates a Gaussian log-likelihood
\begin{equation}
  \ln\mathcal{L} = -\frac{1}{2}\sum_{k}
  \left(\frac{y_{k}^{\mathrm{obs}} - y_{k}^{\mathrm{model}}}
  {\sigma_{k}}\right)^{2},
  \label{eq:loglike}
\end{equation}
where $y_{k}^{\mathrm{obs}}$ and $\sigma_{k}$ are the observed value
and uncertainty of the $k$th constraint, respectively, and
$y_{k}^{\mathrm{model}}$ is the corresponding \texttt{VPLanet} output.
The two constraints used here are the bolometric luminosity
$L_{\mathrm{bol}}$ and the XUV-to-bolometric luminosity ratio
$\log(L_{\mathrm{XUV}}/L_{\mathrm{bol}})$.

The objective function --- the negative log-posterior, incorporating
priors on stellar mass, $f_{\mathrm{sat}}$, age, and
$\beta_{\mathrm{XUV}}$ --- is minimized via differential evolution
\citep{StornPrice97} as implemented in \texttt{SciPy}
\citep{Virtanen20}. We use the \texttt{best1bin} strategy with a
population size of 3 per dimension, dithered mutation in $[0.5, 1.0]$,
a crossover probability of 0.7, and a convergence tolerance of 0.01,
running for up to 50 generations. In other words, the algorithm stops when either 1)
the standard deviation of the negative log-posterior across all members of the population is less than 1\% of the mean, or 2) 50 generations are completed. In practice, we always achieved convergence within 50 generations.

\subsection{Markov chain Monte Carlo}

We will employ 4 different MCMC packages in our study, which we briefly describe below. Each has strengths and weaknesses, but by comparing their results, we gain confidence in the veracity of the Bayesian inference of the Ribas05 parameters.

The \texttt{emcee} package \citep{ForemanMackey13} implements the
affine-invariant ensemble sampler of \citet{GoodmanWeare10}, in which
a set of ``walkers'' explore the parameter space collectively, using
the positions of complementary walkers to propose new steps. Because
the proposal distribution adapts to the covariance structure of the
target, the algorithm is insensitive to linear degeneracies among
parameters and requires no hand-tuning of a proposed covariance
matrix. For computing Ribas05 posteriors, we initialize 100 walkers in a tight ball around the MAP estimate and run each walker for $2\times10^{4}$ steps, discarding autocorrelated
samples until the effective sample size exceeds $10^{4}$.  Note that we also use  \texttt{emcee} to infer the \citet{Davenport19} FFD model parameters except without the MAP initialization.

The \texttt{dynesty} algorithm \citep{Speagle20} implements dynamic nested sampling
\citep{Skilling04,Higson19}, an algorithm that simultaneously
estimates the Bayesian evidence $\mathcal{Z}$ and the posterior. Live
points are drawn from the prior, and at each iteration the
lowest-likelihood point is replaced by a new sample drawn from within
an ``iso-likelihood'' contour. We use multi-ellipsoidal bounding with
random slice sampling \citep{Handley15a,Handley15b} and 50 initial
live points, adding live points dynamically until the estimated
evidence error satisfies $\Delta\ln\mathcal{Z} < 1$.

The \texttt{PyMultiNest} code \citep{Buchner14} provides a Python interface to
the \texttt{MultiNest} library \citep{Feroz09,Feroz19}, which
implements a multimodal nested sampling algorithm. The \texttt{MultiNest} algorithm
partitions the prior volume into multiple ellipsoidal clusters,
enabling efficient sampling of posteriors with well-separated modes.
We run \texttt{PyMultiNest} with 500 live points, a sampling efficiency of 0.8, and
terminate the parameter space exploration when the remaining evidence contribution falls below a
tolerance of $\Delta\ln\mathcal{Z} = 0.5$.

Finally, \texttt{UltraNest} \citep{Buchner21} implements a
reactively-allocated nested sampling algorithm that adjusts the number
of live points during the run to maintain a target accuracy in both
the evidence and the posterior. The algorithm uses the \texttt{MLFriends} (maximum likelihood friends)
region \citep{Buchner19} to draw new live points, which provides
robust performance in high-dimensional or multimodal settings without
requiring user-specified bounding geometries. We use a minimum of
500 live points and terminate the run when $\Delta\ln\mathcal{Z} < 0.5$.

\subsection{Bayesian Inference Accelerated by Machine Learning \label{sec:alabi}}

We will infer posteriors for the Ribas05 model with the \alabi (Active Learning for Accurate Bayesian Inference) code \citep{BirkyBarnes26}. This algorithm first trains a Gaussian process (GP) ``surrogate model'' that represents actual \vplanet output from a sparse sampling of \vplanet simulations across parameter space (training data). The surrogate model is a non-parametric representation of \vplanet simulations from which likelihood evaluations can be computed in $\sim$10 ms each, as opposed to a direct \vplanet run that requires $\sim5$ seconds each in the case of stellar evolution. \alabi then exploits the known uncertainties in the GP's predictions to identify regions where a forward model evaluation would provide the most leverage, \ie high likelihood {\it and} high uncertainty \citep{Kandasamy15,WangLi18,Fleming20}.

A key aspect of this approach is the functional representation of the covariance matrix between the model parameters,  often called the kernel function. We considered three kernels for the Ribas05 model: squared exponential, Mat{\' e}rn-3/2, and Mat{\' e}rn-5/2 \citep{Matern86,Rasmussen06}. \alabi provides a framework to identify the best kernel in conjunction with methods to scale the input and output parameters. We found the squared exponential kernel with no input scaling and min-max output scaling performed best (see the Appendix A and \citet{BirkyBarnes26} for more details).

The code then resamples parameter space with additional \vplanet runs and re-trains the GP surrogate model to improve its accuracy (active learning). Each new point is chosen via the Bayesian active learning for posterior estimation  acquisition function \citep{Kandasamy17}. We perform a total of 500 active learning iterations with each sampler. {\it A priori}, we don't know if 500 points is too few or too many to converge to a robust posterior without overfitting, so we follow the advice of \citet{BirkyBarnes26} and examine the percent error and evaluate several complementary processes to confirm that the model converges, see Appendix B.

We inferred the posteriors for the Ribas05 model via the 4 MCMC methods described in the previous subsection using the surrogate model. Below we present the result of a run with 3000 training points and 500 active learning iterations. This approach generated a normalized root mean squared error of 2.4\%, which is effectively the uncertainty in how close the surrogate model's posterior is to the actual posterior.

\subsection{Convergence Modeling\label{sec:vconverge}}

To calculate the distributions of parameters that do not influence the likelihood, \eg the cumulative XUV flux, we follow the approach described in \citet{Gialluca24}. This algorithm, \texttt{vconverge}, first performs a set of \vplanet simulations with initial conditions drawn from the posteriors inferred by \alabi and builds a preliminary PDF of the cumulative XUV flux. Then, we perform batches of additional trials, again drawing from the \alabi posteriors, until the cumulative XUV flux PDF converges. In this case, the initial run consists of 500 trials and batches of 100 trials. We use the Kolmogorov-Smirnov (K-S) test to compare the distributions, with convergence requiring the K-S statistic to remain below 0.004 for all parameters three times in a row. We find that about 20 batches are required to meet this convergence threshold ($\sim2500$ \vplanet simulations per model).

By inferring posteriors of parameters that directly affect the likelihood and then computing the cumulative XUV flux afterwards, this approach saves significant computational time compared to other machine learning algorithms such as convolutional neural networks. Indeed, the convergence process for computing the cumulative XUV flux PDF is actually the computational bottleneck, requiring several times more CPU time to complete than the posterior estimation for all 4 samplers combined. Nonetheless, the  generation of all the results presented in this article requires $\sim100$ CPU hours on modern hardware\footnote{Scripts to generate the results in this article are available at https://github.com/RoryBarnes/GJ1132/XUV.}.

\section{Reanalysis of Kepler Flares\label{sec:kepler}}

After testing multiple flare evolution models \citep{Loyd18,Davenport19,Ilin19,Guenther20,Feinstein20}, we ultimately found the \citet{Davenport19} matched the observations best and applied it to GJ 1132. However, since the original Davenport et al.~model omitted data points that did not show flares when fitting for $a_1, ..., b_3$, it likely overpredicted the FFD. We thus performed a new MCMC analysis with \emcee that included the non-detections in order to generate a more accurate model for GJ 1132. We used the same data set as Davenport et al., which consists of 347 stars and over 485,000 data points.

\begin{figure*}
\centering
    \includegraphics[width=\textwidth]{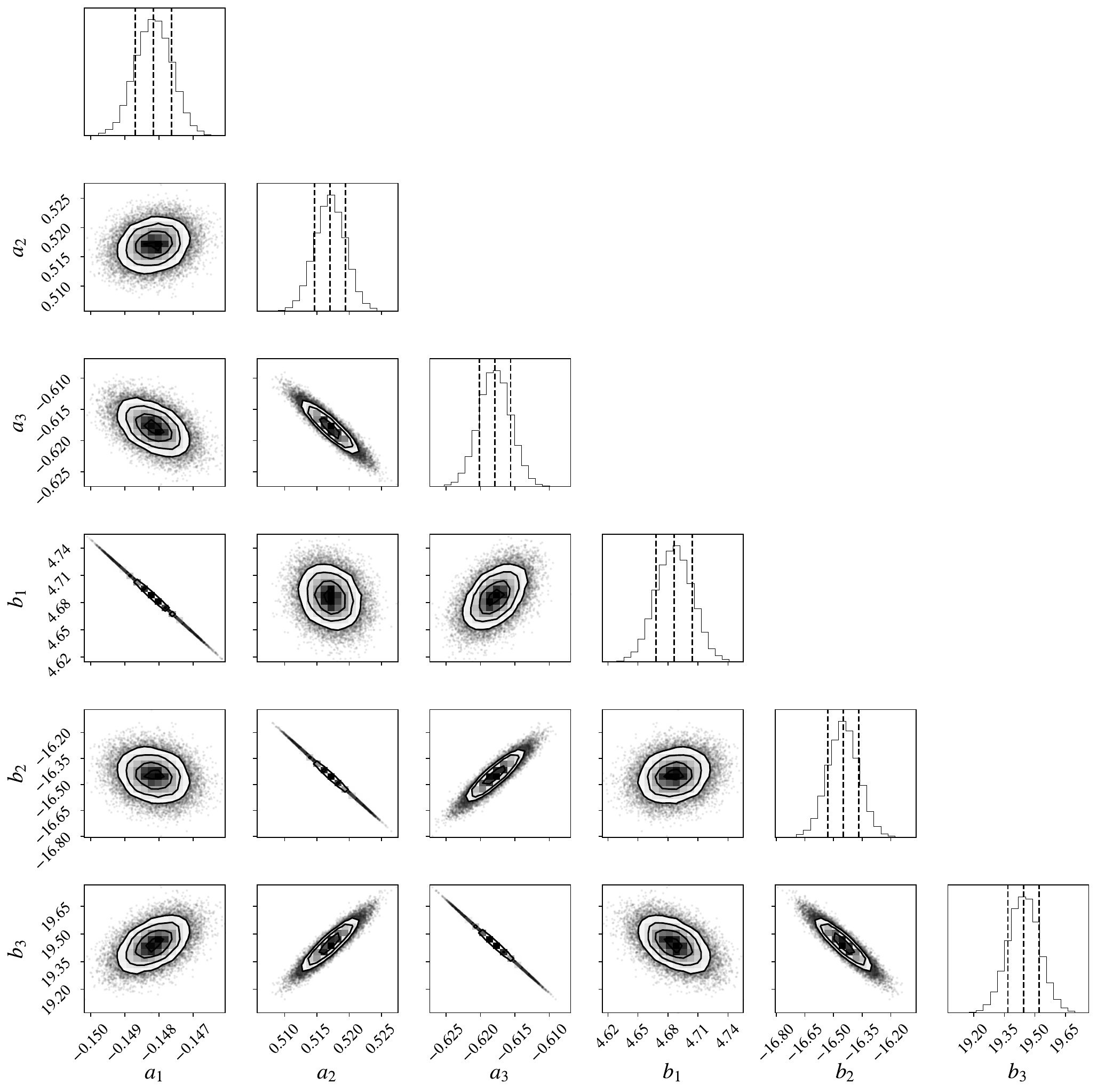}
    \caption{Corner plot for the six-parameter fit for the \citet{Davenport19} FFD model (Eqs.~[\ref{eq:ffd}--\ref{eq:ffd_b}]). The contours denote the 16\%, 50\%, and 84\% confidence intervals, with additional, lower-likelihood points shown by dots. The marginal posteriors along the diagonal include vertical dashed lines that denote the median and 1$\sigma$ uncertainties in the model parameter distributions.}
    \label{fig:ffd_corner}
\end{figure*}

The resulting corner plot is shown in Fig.~\ref{fig:ffd_corner}, which is similar to the results in \citet{Davenport19}, including some tight degeneracies in the covariance matrix, see their Fig.~8. The resulting best fits and 1$\sigma$ uncertainties are shown in Table \ref{tab:ffd}. Note that this result is the global fit, so the parameters apply to any M dwarf, not just GJ 1132.

\begin{table}
\centering
\caption{Fit Parameters for FFD Evolution from \kepler Data}
\begin{tabular}{ll|ll}
\hline\hline
$a_1$ & $-0.148 \pm 0.000529$ & $b_1$ &  $4.69 \pm 0.0180$\\
$a_2$ & $0.517 \pm 0.00244$ & $b_2$ & $-16.45 \pm 0.0824$\\
$a_3$ & $-0.618 \pm 0.00231$ & $b_3$ & $19.446 \pm 0.0779$\\
\hline
\end{tabular}
\label{tab:ffd}
\end{table}

\begin{figure*}
\centering
    \includegraphics[width=\textwidth]{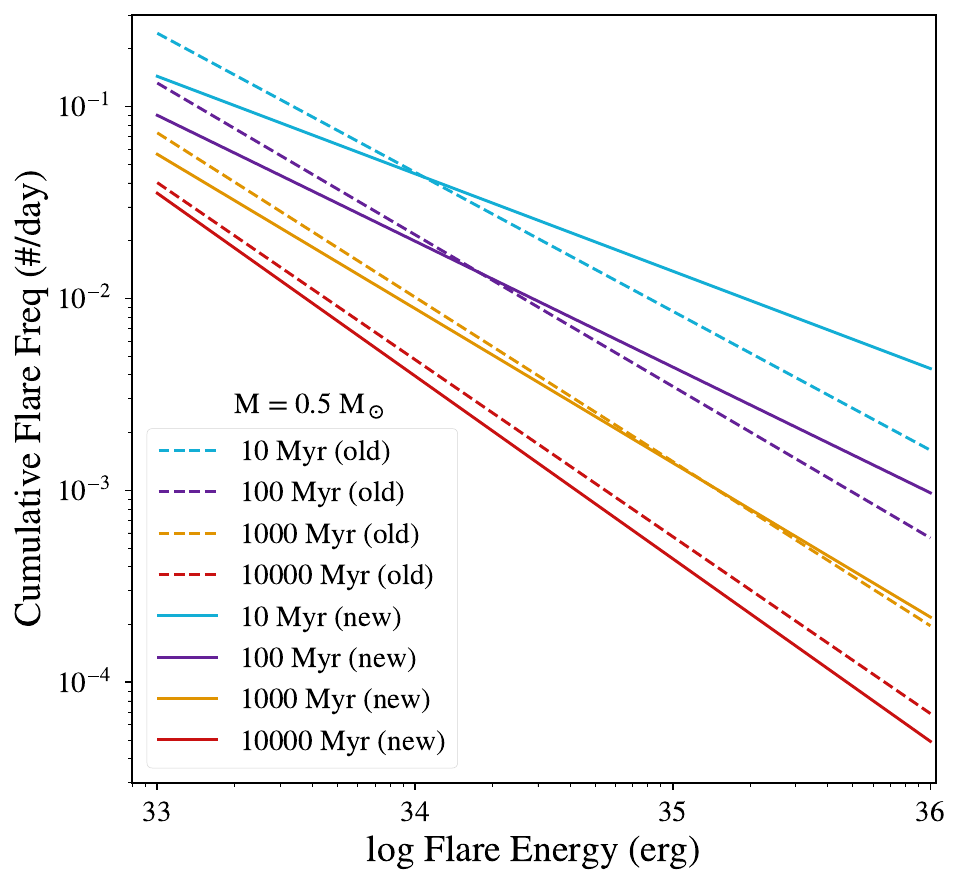}
    \caption{Comparison of the new FFD model (solid line) to the original \citet{Davenport19} FFD model (dashed lines) for 4 different ages of a 0.5 $\msun$ star.}
    \label{fig:ffd_compare}
\end{figure*}

Figure \ref{fig:ffd_compare} compares the original fit to the new one for a 0.5 $\msun$ star, which is a mass that Davenport et al. focused on. We find that the predicted FFDs are similar, but the new fits predict lower overall flare rates as expected. At younger ages, the slopes are flatter, indicating higher energy flares are more common. We stress that, as in Davenport et al., the fit does not match all stars well, but does accurately predict the observed FFDs of many stars, \ie M dwarfs show high intrinsic variance, an observation that will be relevant later. We also tried a three-parameter fit in which $b = -1$ for all cases, but found that it did not perform significantly better according to both the Bayes factor and Bayesian information criterion. We therefore adopt the posteriors for the full model for our modeling of GJ 1132. Although there are degeneracies in the model, our procedure handles them cleanly, see $\S$~\ref{sec:stats}.

\section{Current Flare Rate\label{sec:obsflare}}

FFDs can be difficult to measure for many stars, but the high cadence, long baseline observations of the \kepler and \tess spacecraft have enabled robust flare detections \citep{Davenport19,Guenther20}. Fortunately, GJ 1132 lies in a \tess field, and so we can use the \tess data to obtain an FFD for GJ 1132 today.

We identified stellar flare candidates in the TESS photometry of
GJ~1132 using a semi-automated detection pipeline followed by visual
classification.  The pipeline ingests all available TESS 2-minute
cadence lightcurves from the Science Processing Operations Center
\citep[SPOC;][]{Jenkins2016} via the \texttt{lightkurve} package
\citep{Lightkurve2018}, covering Sectors~9, 10, 36, 63, 90, and~99, for a total of 123.31 days of data.

Each sector's lightcurve is first normalized to unit median flux using
the \texttt{lightkurve} \texttt{normalize} method.  We then remove
low-frequency stellar variability by applying a Savitzky--Golay filter
\citep{Savitzky1964} with a window length of 101~observations
($\approx$3.4~hr at 2-minute cadence).  The \texttt{lightkurve}
implementation iteratively sigma-clips outliers during the fit,
preventing flare signals from biasing the trend removal
\citep{Lightkurve2018}.  The resulting flattened lightcurve isolates
short-duration transient events against a unit baseline.

We estimate the photometric noise in each sector using the Median
Absolute Deviation (MAD), a robust measure of statistical dispersion
that is insensitive to outliers such as flares
\citep{Rousseeuw1993}.  For a sample of $N$ flux measurements
$\{F_i\}$, the MAD is defined as
\begin{equation}\label{eq:mad}
  \mathrm{MAD} = \mathrm{median}\bigl(|F_i - \widetilde{F}|\bigr),
\end{equation}
where $\widetilde{F} = \mathrm{median}(F_i)$.  The MAD is related to
the standard deviation of a Gaussian distribution by the consistency
constant $c = 1/\Phi^{-1}(3/4) \approx 1.4826$, where $\Phi^{-1}$ is
the quantile function of the standard normal distribution.  The
MAD-based robust sigma is therefore
\begin{equation}\label{eq:robust_sigma}
  \hat{\sigma} = 1.4826 \times \mathrm{MAD}.
\end{equation}
Unlike the sample standard deviation, $\hat{\sigma}$ has a breakdown
point of 50\%, meaning that up to half the data can be arbitrarily
corrupted without affecting the estimate \citep{Rousseeuw1993}.  This
property makes it well-suited for flare detection, where bright
transient events would inflate a conventional $\sigma$ estimate and
suppress the detection of weaker flares.

We flag all cadences in the flattened lightcurve whose flux exceeds
$\widetilde{F} + n_\sigma\,\hat{\sigma}$, where $n_\sigma$ is the
detection threshold.  We adopt $n_\sigma = 2.5$ to ensure sensitivity
to low-amplitude flares at the expense of a higher false-positive rate
that is subsequently filtered by visual inspection.  Individual
above-threshold cadences are grouped into contiguous clusters, allowing
a gap tolerance of up to 2 cadences ($\approx$4~min) to bridge
single-point dropouts within a flare event.  Clusters containing fewer
than 2 consecutive above-threshold values are
discarded since isolated single-point excursions are more likely due to
cosmic ray hits or instrumental artifacts than to stellar flares
\citep{Ilin2021}.

Stellar flares tend to exhibit a characteristic fast-rise, exponential-decay
evolution, but more complex morphologies are also possible \citep{Davenport2014}.  While the impulsive rise is
typically captured by the above-threshold cadences, the exponential
decay tail often extends well below the detection threshold.  To
capture the full flare energy, we extend each candidate window forward
in time from the last above-threshold cadence until the flux returns
to within $1\,\hat{\sigma}$ of the median for at least 3 consecutive
observations ($\approx$6~min).  This criterion balances the need to
capture the decay phase against the risk of absorbing unrelated
variability and is ultimately verified via the visual inspection step.

For each candidate, we record 1) the TESS Barycentric
Julian Date of maximum flux in the candidate window, 2) the peak flux
expressed in units of $\hat{\sigma}$ above the sector median, 3) the time interval from the first
above-threshold cadence to the end of the decay tail, in minutes, and 4)  the
time-integrated relative flux excess,
\begin{equation}\label{eq:equiv_dur}
  \mathrm{ED} = \int_{t_\mathrm{start}}^{t_\mathrm{stop}}
    \bigl(F(t)/F_\mathrm{quiescent} - 1\bigr)\,dt,
\end{equation}
with units of seconds \citep{Hunt-Walker2012}.  The equivalent
duration (ED) is proportional to the flare energy and independent of
distance, making it a natural unit for constructing flare frequency
distributions \citep{Hawley2014}.  We evaluate the integral
numerically using the trapezoidal rule on the normalized
(unflattened) lightcurve, as the flattening process can distort
the absolute flux scale of long-duration events.

\begin{figure*}
\centering
    \includegraphics[width=\textwidth]{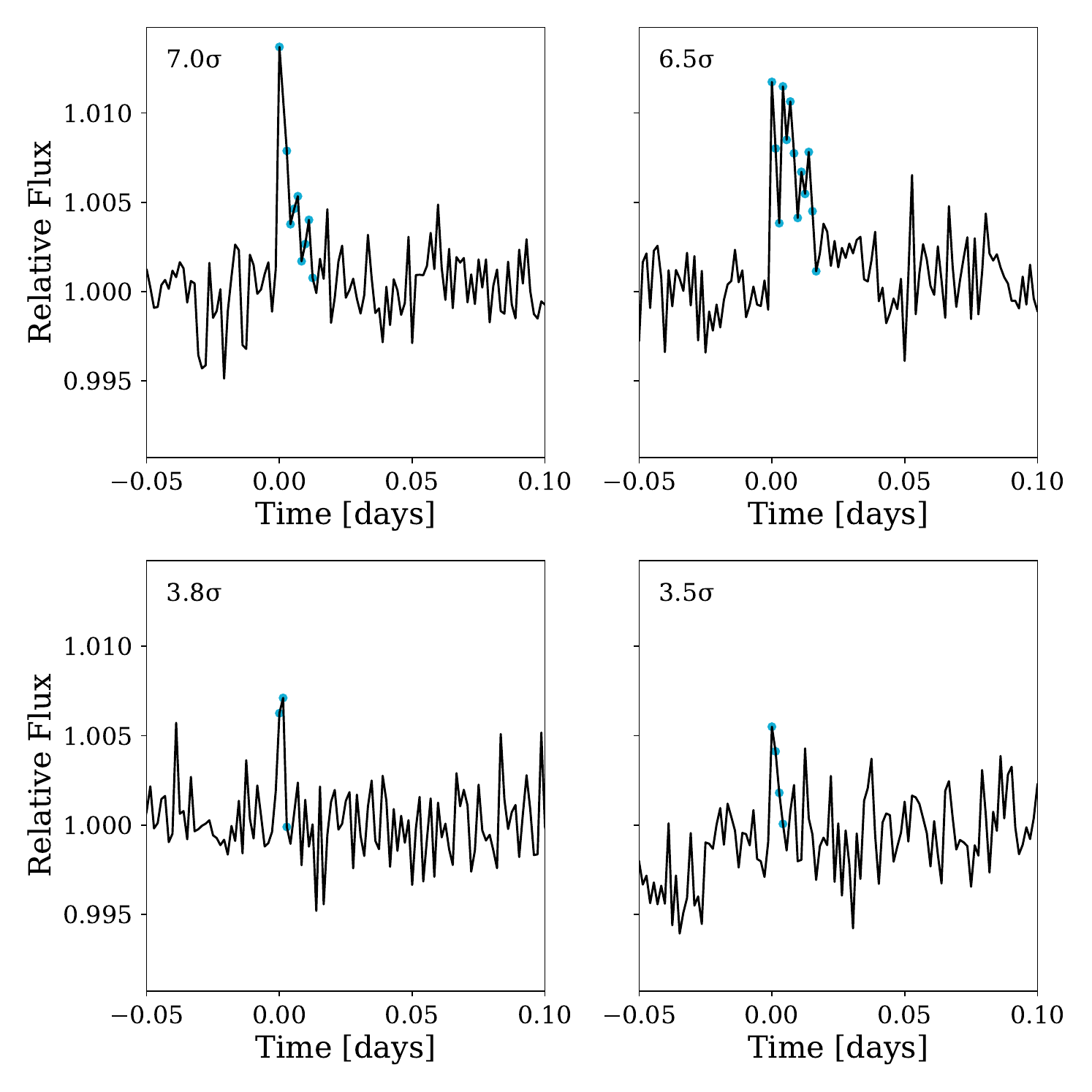}
    \caption{Four flare candidates of GJ 1132 observed by \tess. The blue points indicate times that are labeled as flaring by our automated pipeline and used to compute the total energy of the flare. Note also that the baseline flux varies with an amplitude of about 1\% suggesting that the star has a natural variability at this level on $\sim$2-minute timescales.}
    \label{fig:flareex}
\end{figure*}

Each candidate is then inspected visually over a range
spanning $[-0.05, +0.10]$~days ($\approx$[$-72$, $+144$]~min) relative
to the peak time.  In total, 16 events met the automated detection threshold, but only 4 were deemed to be flares, as shown in Fig.~\ref{fig:flareex}, and none exceeded a 1.5\% increase in brightness. Two of these flares are unambiguous via both the statistical metric as well as visual inspection. The other two, however, should be considered marginal detections, with normalized peak fluxes that reach just 3.5--4$\sigma$ above the mean flux.

We fit Eq.~(\ref{eq:ffd}) to the four TESS flares using
weighted least squares, obtaining a slope of $a = -0.45 \pm 0.72$ and a $y$-intercept of $b = 12.5 \pm 22.7$.  Because the four
flares span barely one order of magnitude in energy ($31.1 \lesssim
\log_{10}(E) \lesssim 32.2$), where $E$ is the total flare energy in ergs, the slope and intercept are almost
perfectly anti-correlated, with a correlation coefficient  of -0.9999. This near-perfect anti-correlation
means the joint constraint occupies a very narrow ellipse in $(a,
b)$ space, even though the marginal uncertainties on each
parameter are large.

\begin{figure*}
\centering
    \includegraphics[width=\textwidth]{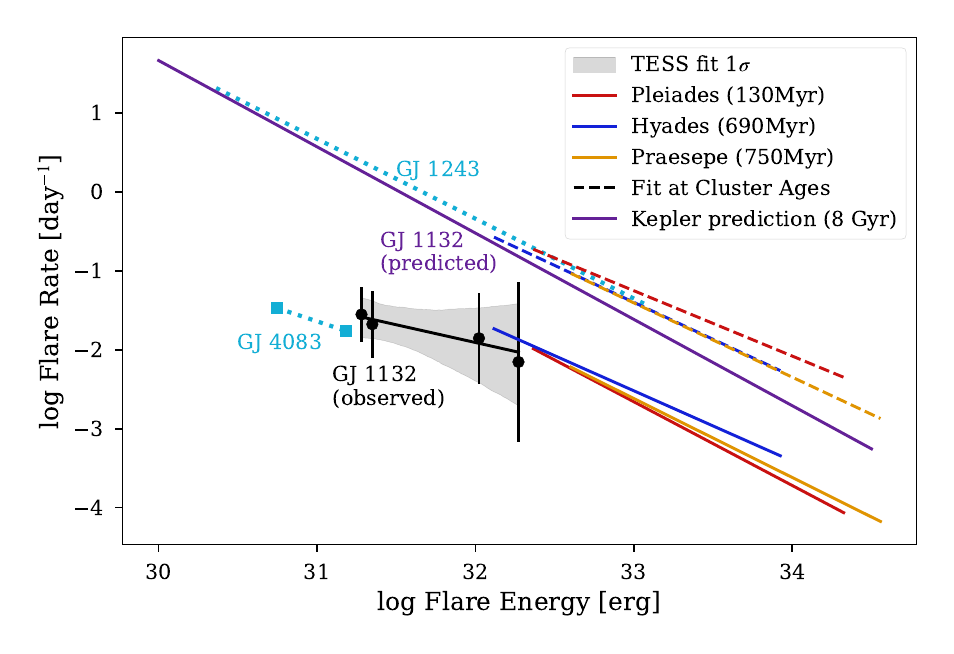}
    \caption{GJ 1132's FFD from \tess data (black points), a linear regression fit to the data (black line), and the 1$\sigma$ uncertainty in the fit (grey shaded region). Additional FFDs are included for reference, including the average of several clusters (dashed red, dark blue, and orange lines) as well as the prediction of the flare model for GJ 1132 with an age of 8 Gyr (purple lines). The FFDs of two M dwarfs observed with \kepler, GJ1243 (M4 spectral type; age $\sim 1$ Gyr \citep{Davenport20}) and GJ 4083 (M3 spectral type; age poorly constrained), are shown with dotted pale blue lines.}
    \label{fig:ffd}
\end{figure*}

We show the derived FFD in Figure \ref{fig:ffd} along with the FFD of several other stars for reference. Although the uncertainty in the current FFD is quite large, we find it is 7$\sigma$ away from the predicted slope from the \kepler reanalysis for a 0.2 solar mass star with an age of 8 Gyr. We caution that 123 days is not nearly as long as the 4 year baseline of the \kepler mission and 4 flares does not provide strong constraints on the current FFD, so this discrepancy may not be meaningful. Nevertheless, the low apparent flaring rate suggests GJ 1132 is an old star, which is consistent with its large rotation period of $\sim125$ days.

Ultimately, we choose the theoretical model from \citet{Davenport19} to model the evolution of $L_{XUV}^{flares}(t)$, independent of the observed FFD. The small number of flares does not offer a robust calibration against the model, nor does it provide a strong constraint on age. In Section \ref{sec:results}, we include the contribution of flares according to the Davenport et al.~model, but given the analysis presented in this section, the predictions for the XUV luminosity from flares should probably be viewed as an upper limit. In terms of planet b's distance from the cosmic shoreline, however, we will see that the role of flares is likely to be inconsequential.

\section{Current Quiescent XUV \label{sec:quiescent}Luminosity}

\begin{figure*}
    \includegraphics[width=0.47\textwidth]{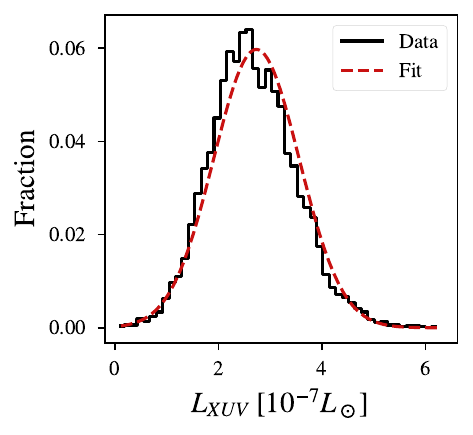}
    \includegraphics[width=0.47\textwidth]{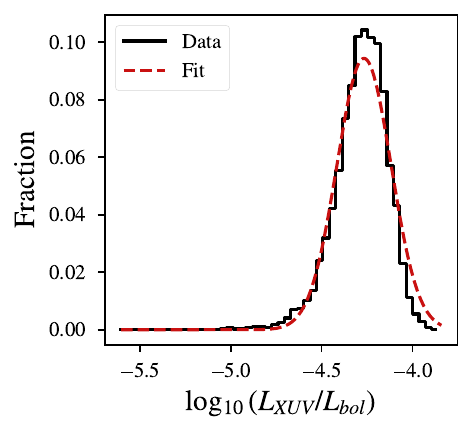}
    \caption{{\it Left:} The distribution of quiescent XUV luminosity of GJ 1132. {\it Right:} The distribution of the ratio of the quiescent XUV-to-bolometric luminosity of GJ 1132.}
    \label{fig:quiescentXUV}
\end{figure*}

Next we turn to calculating the current quiescent XUV luminosity of the host star, which enables the inference of the \citet{Ribas05} model parameters. As shown in Table~\ref{tab:starprops}, only the X-ray luminosity has been directly measured, so we must rely on a scaling relationship to calculate the XUV luminosity, $L_{EUV}$. For this calculation, we use the \citet{SanzForcada25} model:
\begin{equation}
    \log(L_{EUV}) = (0.821 \pm 0.041)C_X + (28.16 \pm 0.05),
    \label{eq:leuv}
\end{equation}
where
\begin{equation}
    C_X = \log(L_X) - 27.44,
\end{equation}
when all luminosities are expressed in units of ergs/s. We first build the $L_{XUV}$ distribution via $10^4$ draws from the observed $L_X$ distribution, and then compute an $L_{EUV}$ value from it, including the inherent uncertainties in the \citet{SanzForcada25} model. The result is shown in the left panel of Fig.~\ref{fig:quiescentXUV}, where the best fit and uncertainty are $L_{XUV} = (2.74 \pm 0.83) \times 10^{-7} L_\odot$. Figure \ref{fig:quiescentXUV} shows that the uncertainties are also well-represented by a Gaussian.

While the absolute value of the XUV luminosity is valuable, the \citet{Ribas05} model requires the ratio of $L_{XUV}/L_{bol}$, see Eq.~(\ref{eq:lxuv}). We thus created this distribution via $10^4$ independent draws from the normally distributed $L_{XUV}$ distribution and the asymmetric uncertainties in $L_{bol}$ (see Table \ref{tab:starprops}) and find $\log(L_{XUV}/L_{bol}) = -4.26 \pm 0.15$. The resulting distribution is shown in the right panel of Fig.~\ref{fig:quiescentXUV}, which shows some skew due to the asymmetric uncertainties in $L_{bol}$, especially at low values. Nonetheless, the uncertainties are close enough to a normal distribution that we adopt the symmetric Gaussian to represent the quiescent XUV luminosity, which we then use to constrain the Ribas05 model in the next section.

\section{Stellar XUV Evolution\label{sec:results}}

In this section we present the results of our stellar modeling, including total XUV from quiescence and flares. We first use gyrochronology to generate an age PDF, which we then use as a prior to infer the posteriors of the Ribas05 model. Since the EMD model is analytic, we do not need to infer any of its parameter's PDFs via Bayesian methods -- its properties can be estimated via standard Monte Carlo techniques.

\begin{figure}
    \includegraphics{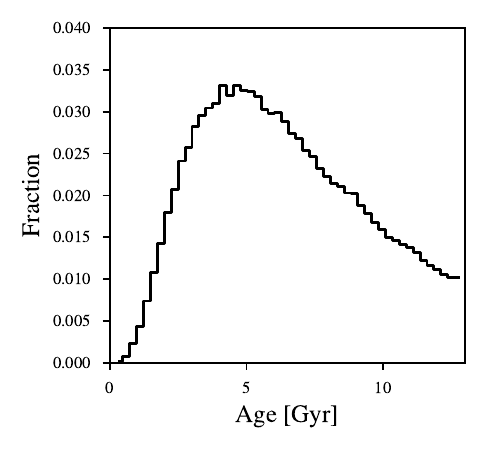}
    \caption{Distribution of GJ 1132 ages predicted by the \cite{Engle24} model, truncated at 13 Gyr.}
    \label{fig:EngleAges}
\end{figure}

Figure \ref{fig:EngleAges} shows the distribution of ages predicted by the \cite{Engle24} model based on the observed rotation period. The \citet{Engle24} model relies on measuring the rotation periods of M dwarfs whose age can be inferred by association, e.g., M dwarfs in clusters, with a white dwarf companions, etc., so is a reasonable method to constrain M dwarf ages. In log space, the distribution is well-described by a Gaussian distribution: $\tau = 0.76 \pm 0.22$, which corresponds to a mean age of $\sim$6 Gyr. Despite the long tail to old ages, the mode of the distribution lies at just under 5 Gyr. Note that we assume the star cannot be older than 13 Gyr.

Turning to the Ribas05 model, we use the $L_{bol}$ and $L_{XUV}/L_{bol}$ constraints to infer posteriors for its model parameters. Table \ref{tab:ribas} lists the priors, expressed as Gaussians, and the search bounds for the model parameters. The $t_{sat}$ prior is uninformative, while the age prior is not shown because it is drawn from the EMD gyrochronology model. We then followed the procedure described in $\S$~\ref{sec:stats}, first computing the maximum likelihood, then using \alabi to infer the posteriors with MCMC sampling coming from \emcee, \dynesty, \pymultinest, and \ultranest. More details on this procedure are presented in the appendices.

Figure~\ref{fig:XUVPosteriors} shows the posteriors for the Ribas05 XUV model for the 4 sampling techniques. All 4 show similar posteriors, suggesting the priors and data do generate a reasonable posterior. Although we use an uninformative prior for $t_{sat}$, we find it is very likely to be less than 3 Gyr, and probably less than 1 Gyr. This result is driven largely by the relatively low current $L_{XUV}$ value (see Fig.~\ref{fig:quiescentXUV}), which suggests a short saturation time. This analysis also strongly favors an older age for the system, with significant likelihood out to our upper bound of 13 Gyr. Note that the posterior is wider than the prior at older ages.

\begin{table*}
\centering
\caption{Analytic Priors and Bounds Used for Constraining the \citet{Ribas05} Model Parameters}
\begin{tabular}{l|cc|cc}
\hline\hline
Parameter & Mean & Standard Deviation & Lower Bound & Upper Bound\\
\hline
Mass [$M_\odot$] & 0.181 & 0.019 & 0.1 & 0.3\\
log($f_{sat}$) [Gyr] & -2.92 & 0.26 & -4.0 & -1.0\\
$t_{sat}$ [Gyr] & - & - & 0.1 & 5\\
$\beta_{XUV}$ & -1.18 & 0.31 & -2.0 & 0.0\\
\hline
\end{tabular}
\label{tab:ribas}
\end{table*}

\begin{figure*}
    \includegraphics[width=\textwidth]{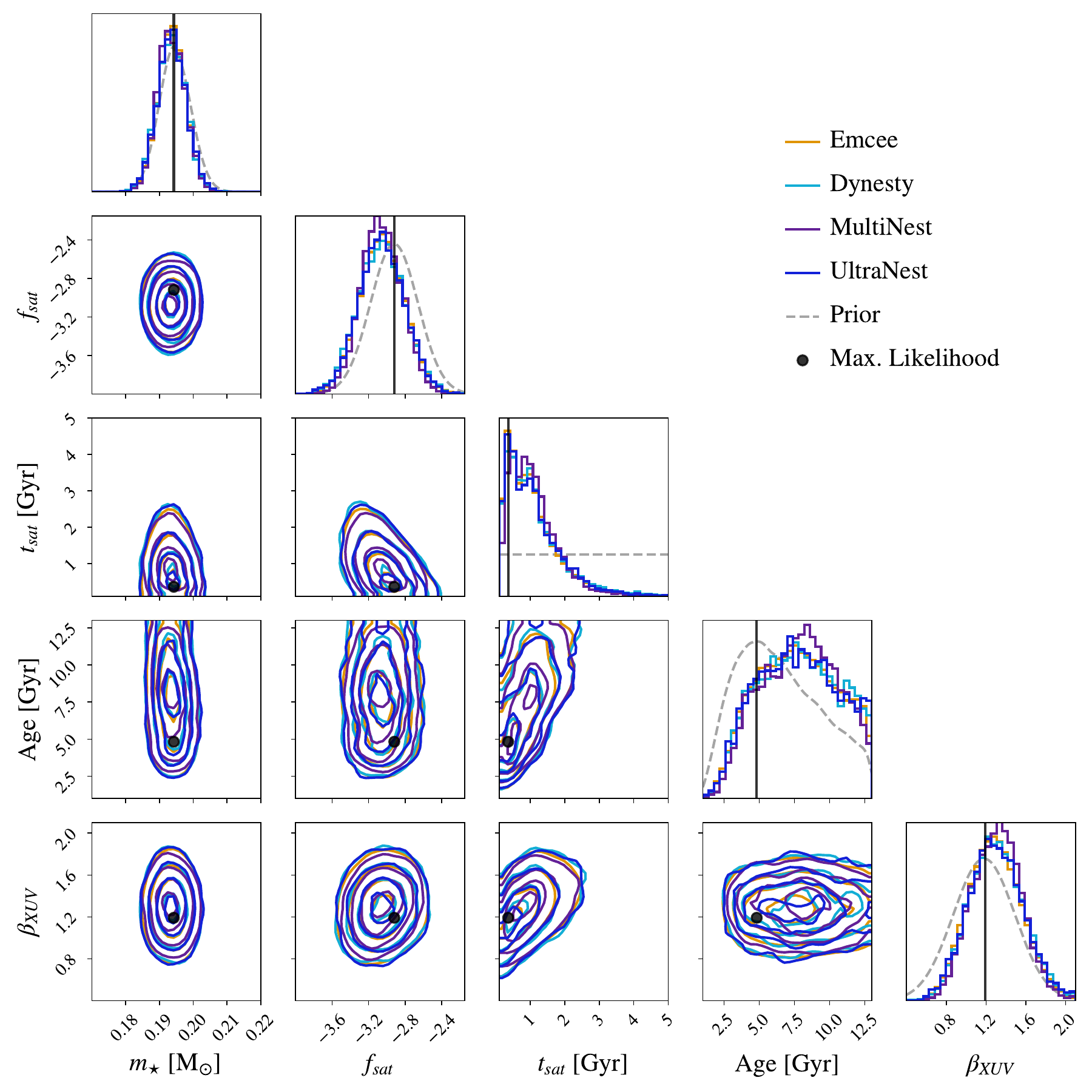}
    \caption{Corner plot for the \cite{Ribas05} model parameters. Pale blue curves are from \texttt{dynesty}, orange from \texttt{emcee}, purple from \texttt{MultiNest}, and dark blue from \texttt{UltraNest}. The grey curves are priors and the black dots and vertical lines in the marginal posteriors denote the maximum likelihood.}
    \label{fig:XUVPosteriors}
\end{figure*}

\begin{figure*}
    \includegraphics[width=\textwidth]{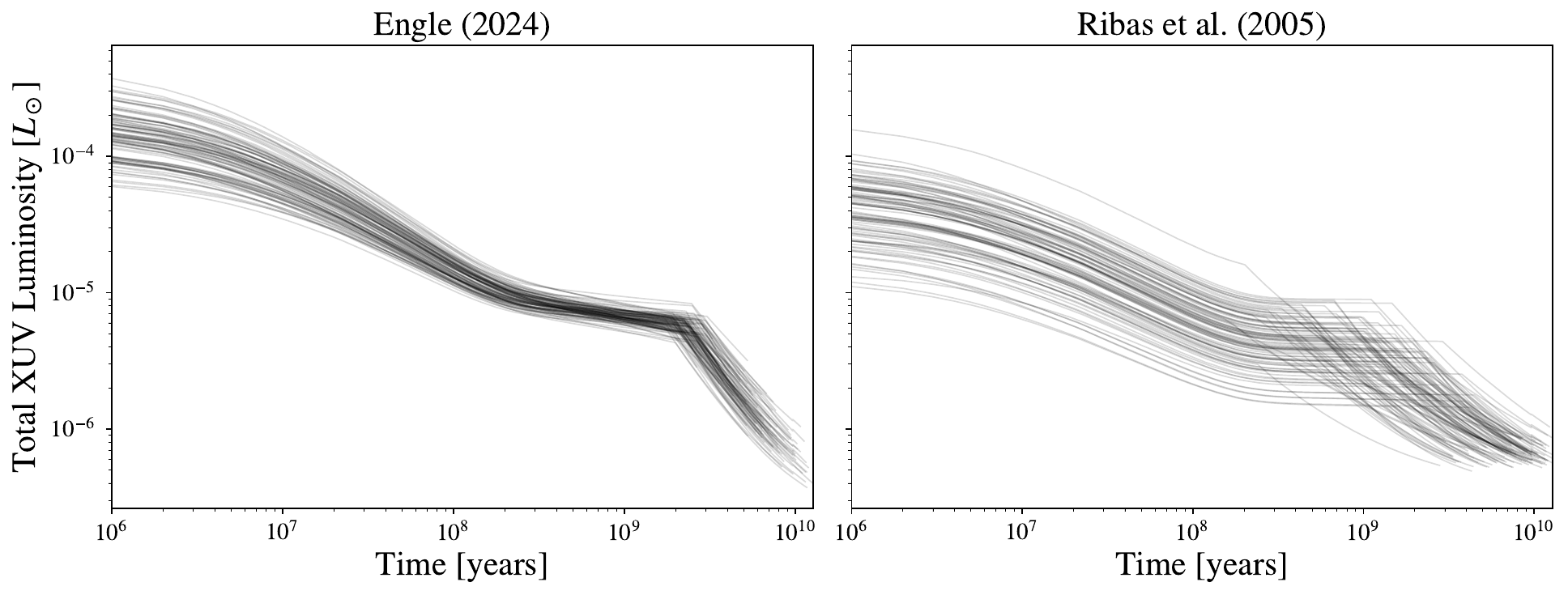}
    \caption{Example XUV evolutions for GJ 1132. Both panels include flares, but the left panel assumes the quiescent model of \citet{Engle24} and the right assumes the model from \citet{Ribas05}.}
    \label{fig:example}
\end{figure*}

Figure \ref{fig:example} shows 100 representative examples for each quiescent model plus flares. The discontinuities at larger ages represent the end of the saturation period. This figure shows that the \citet{Engle24} model predicts larger XUV fluxes at earlier ages, but a more rapid fall-off over time compared to \citet{Ribas05}, especially after saturation. Note as well that the the EMD curves are more tightly clustered than the Ribas05 trajectories.

\section{Cumulative XUV Flux on GJ 1132 b}

With the distributions of XUV luminosity over time computed, we turn to its potential impact on planet b. Before proceeding, we must first recognize that the cosmic shoreline formulation relies on a comparison between the cumulative XUV on an object in comparison to that of Earth, so we must compute that value. Using the \cite{Ribas05} model, which is preferable here as it was derived from solar twin data, with a saturation time of 100 Myr and a saturation fraction of $10^{-3}$, the cumulative quiescent flux the Earth has received is $F_{XUV,\oplus}^{Cumulative} = 9.8 \times 10^{15}$ W. We assume that Earth's distance from the relatively quiet Sun implies that the solar flare contribution to Earth's cumulative XUV flux is negligible.

\begin{figure*}
    \includegraphics[width=0.5\textwidth]{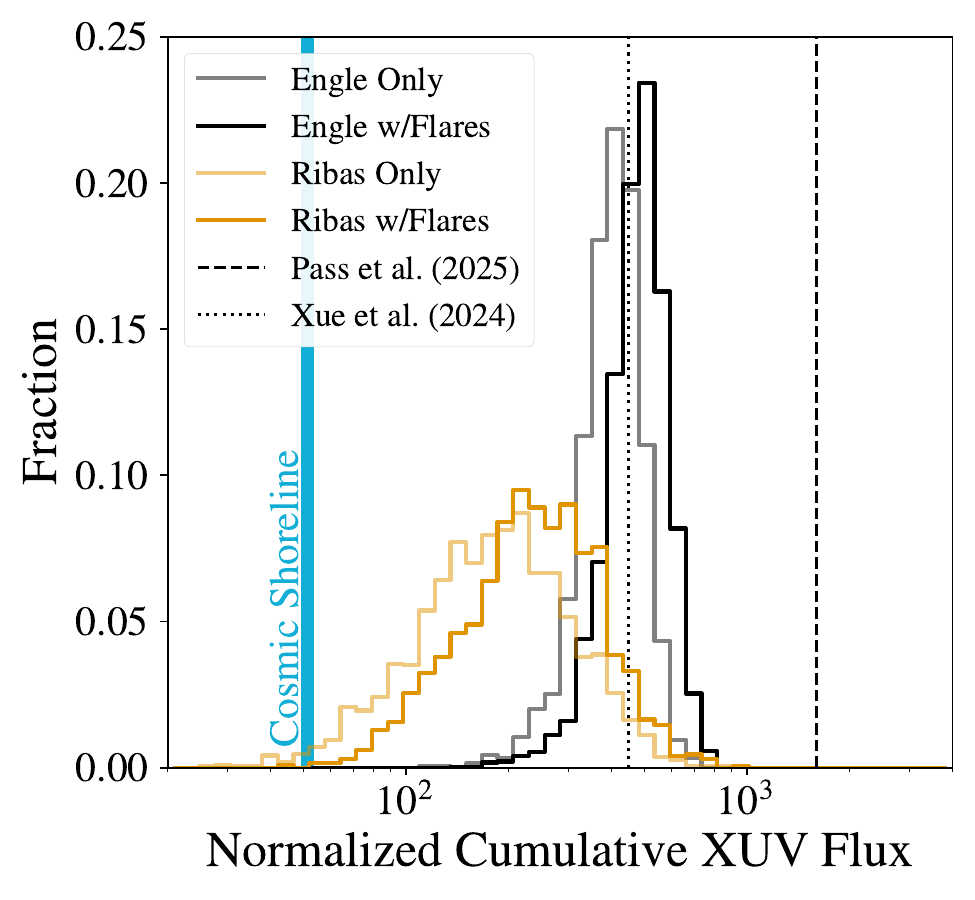}
    \includegraphics[width=0.5\textwidth]{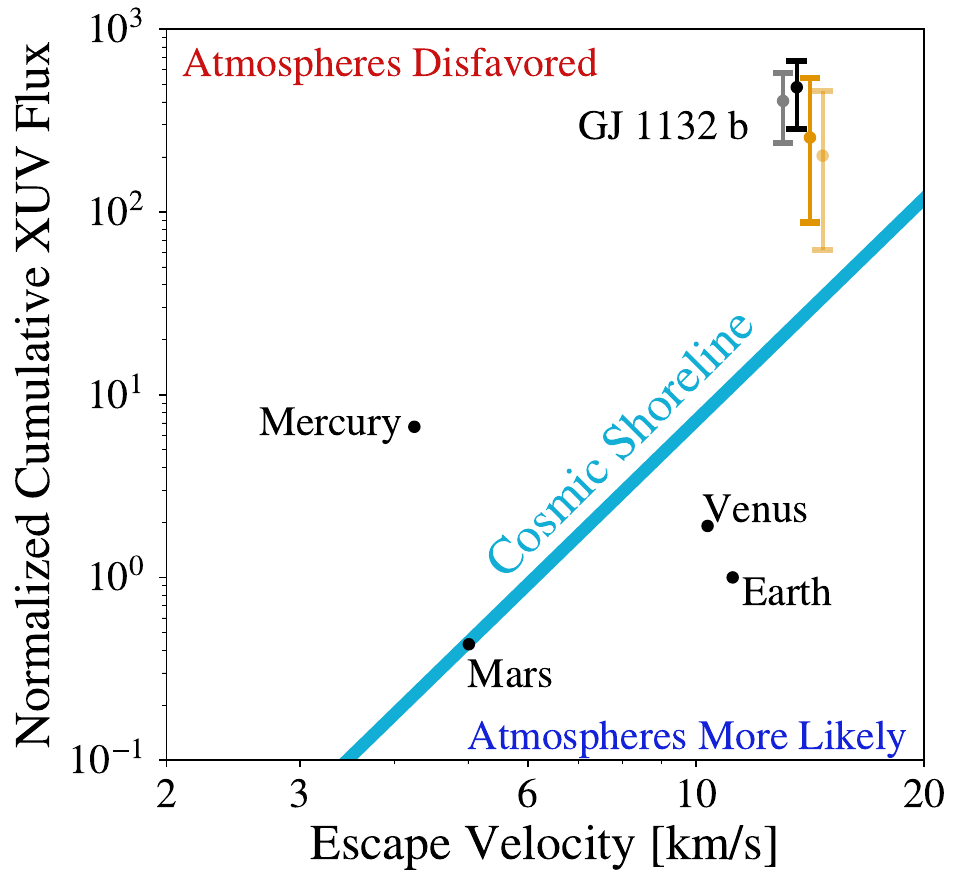}
    \caption{{\it Left:} Histograms of cumulative XUV fluxes planet b has received for different assumptions. The recently proposed values from Xue24 and \citet{Pass25} are also shown for reference. {\it Right:} The distribution of XUV fluxes for GJ 1132 b in relation to the ``cosmic shoreline'' \citep{ZahnleCatling17}. The uncertainties shown are two sigma. The GJ 1132 b points are offset slightly in the $x$-axis to improve readability. The terrestrial planets of the Solar System are shown for reference.}
    \label{fig:shoreline}
\end{figure*}

Figure \ref{fig:shoreline} shows these XUV ranges in relation to the cosmic shoreline  \citep{ZahnleCatling17}. The left panel shows the PDFs for 4 different models in comparison the critical flux for a planet with the mass and radius of GJ 1132 b. Among all our trials, only the Ribas05 model without flares includes cases in which the planet lies on the atmosphere-expected side, and even then, $<$1\% meet the criterion. Figure \ref{fig:shoreline} also shows the position of GJ 1132 b in the standard cosmic shoreline parameter space with the 95\% confidence intervals for the four models. Table \ref{tab:results} shows the resulting distribution of the normalized cumulative XUV flux on planet b, revealing that flares contribute about 20\% of the total cumulative XUV flux, consistent with the results of \citet{Pass25}. Thus, it seems very likely that, even when accounting for uncertainties, the cosmic shoreline conjecture strongly favors an atmosphere-free world.

\begin{table}
\centering
\caption{Cumulative XUV Flux on GJ 1132 b (Earth Units) }
\begin{tabular}{lcc}
\hline\hline
Model & Mean & 95\% Confidence Interval\\
\hline
EMD Only & 402 & [241, 571]\\
EMD w/Flares & 484 & [291, 671]\\
Ribas Only & 204 & [62, 477]\\
Ribas w/Flares & 257 & [87, 523]\\
\hline
\end{tabular}
\label{tab:results}
\end{table}

\section{Discussion\label{sec:discussion}}

We have performed a statistical analysis of the historical XUV luminosity of GJ 1132, a nearby planet-hosting star whose innermost planetary companion is the target of ongoing JWST campaigns to detect and characterize a putative atmosphere. We are able to constrain its quiescent and flaring XUV luminosity over time and find that planet b has likely intercepted over 200x more XUV photons than Earth. Combined with the known mass and radius, we find that the cosmic shoreline hypothesis \citep{ZahnleCatling17} predicts the planet will be devoid of an atmosphere, a prediction currently being checked via \jwst observations.

This work constitutes the first rigorous treatment of the uncertainties of the cumulative XUV flux for a terrestrial \jwst target, although note that \citet{Pass25} do qualitatively discuss uncertainties in the cumulative XUV of these planets. Our results reveal that the uncertainties in this value predict a range that spans a factor of 2.4 (EMD only) to 7.7 (Ribas05 only). Given that the EMD model is both calibrated to M dwarfs and produces a tighter probability distribution, we argue that it is the better measure of cumulative XUV flux for this system. For this model, even the lowest permitted value for cumulative XUV flux is a factor of 2 above the cosmic shoreline.

These results are significantly different than the recent predictions of \citet{Pass25}, who derived a different model for the historical XUV luminosity of mid- to late-M dwarfs based on their activity lifetimes. Their model does not include uncertainty estimation, but they do report that a quiescent + flare model for GJ 1132 b predicts a cumulative XUV flux on planet b of 1600 times that of Earth. The comparable model in our analysis is the EMD w/Flares model, which has a best fit value near 500, a factor of 3 lower, but also $\sim$10$\sigma$ from our result. Since their model does not include uncertainties, we cannot rigorously compare the results, but it seems there is a large discrepancy between the two results that needs to be resolved.

\begin{figure*}
    \includegraphics[width=\textwidth]{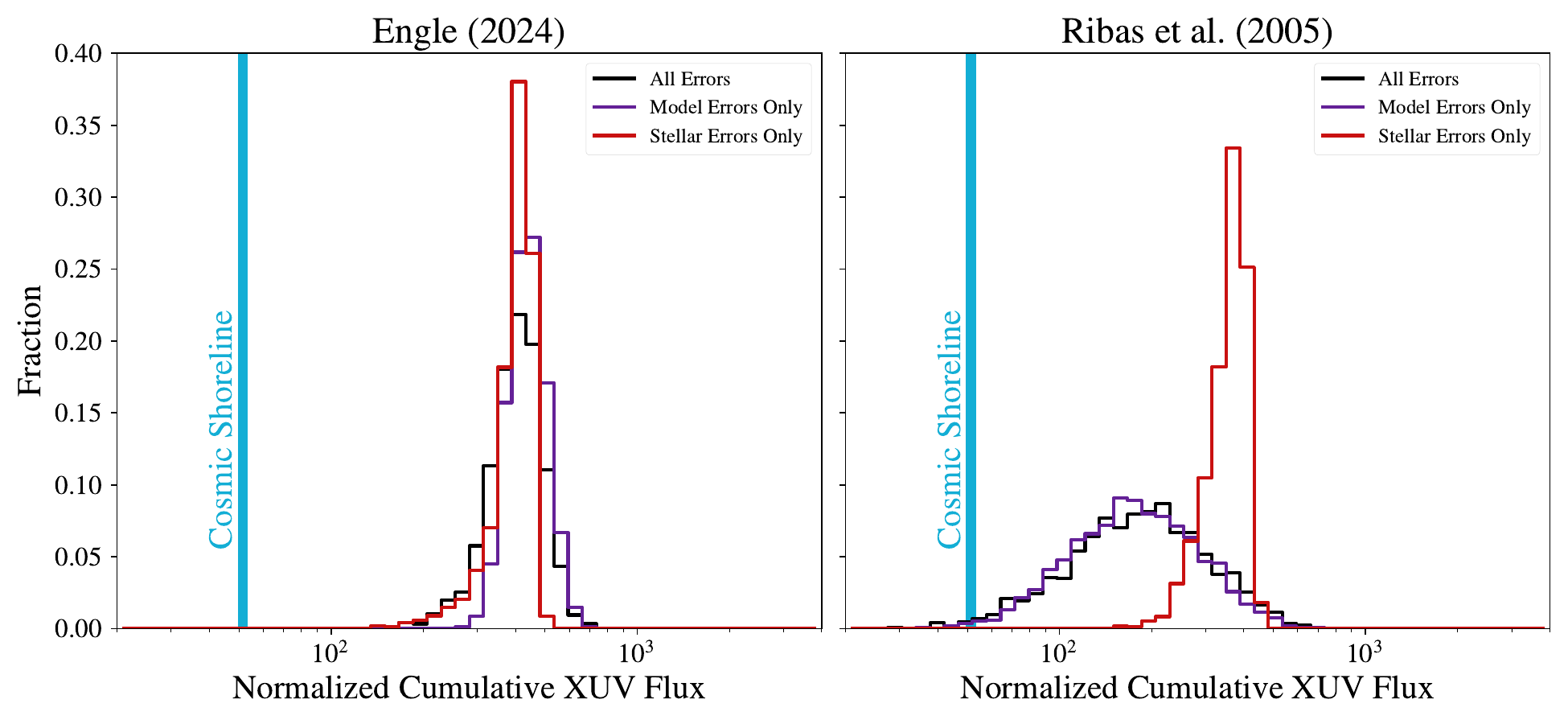}
    \caption{Comparison of cumulative XUV flux distributions for the \citet{Engle24} model (left) and \citet{Ribas05} model (right), but highlighting the different error sources. The red histograms only included uncertainties in the fundamental stellar parameters (mass and age); the purple histograms only included uncertainties in the model parameters; the black histogram includes both. The cosmic shoreline is shown in pale blue for reference.}
    \label{fig:ErrorCompare}
\end{figure*}

A detailed comparison of the Ribas05 and EMD XUV models reveals an interesting difference between the two in terms of the dominant source of uncertainty. In Fig.~\ref{fig:ErrorCompare}, we plot the cumulative XUV flux for the two models with all uncertainties (black), model errors only (purple; stellar parameters held fixed at the best-fit and an age of 8 Gyr), and stellar errors only (red; model parameters held fixed at their best fit value). For the Ribas05 model errors, we used the \dynesty posterior shown in Fig.~\ref{fig:XUVPosteriors}, including age. Flaring is not included in these distributions. For the newer EMD model, the distributions are very similar, suggesting that neither the stellar or model uncertainties are dominating the overall error budget. For the Ribas05 model, however, the uncertainties in the model parameters clearly dominate. Thus, while the two models predict about a factor of 2 difference between the cumulative XUV flux, the \citet{Engle24} model has clearly improved the precision of XUV modeling.

The addition of flares to the total cumulative XUV flux represents a conceptual change to the cosmic shoreline model, so may not be appropriate to include. The original formulation is based solely on the quiescent XUV flux of the Sun, ignoring relatively infrequent flares that are likely fairly diluted by the time they reach the terrestrial planets of the Solar System. Thus, their inclusion in the analysis for short-period planets orbiting M dwarfs appears warranted, even if they only marginally increase the cumulative XUV flux. In this case, the flare observations proved too sparse to provide a robust constraint on the flaring history, but we also see that their inclusion does not qualitatively affect our results.

The revised Kepler flare model predicts lower overall flare rates, see Fig.~\ref{fig:ffd}, but still predicts a significantly steeper and higher FFD for GJ 1132 than we found from \tess data, see Fig.~\ref{fig:ffd_compare}. However, it is important to note that we found that many \kepler flares possess a similar morphology to GJ 1132. On the other hand, \kepler's better sensitivity may enable more flare detections than \tess, so the flaring rate we obtained may be lower than an equivalent observation with \kepler. Future work should continue to refine our understanding of M dwarf flare rates over time, bearing in mind that multiple evolutionary pathways may be possible for these stars. In our modeling, we find that 20\% of the cumulative XUV flux comes from flares, but given the low rate of flaring today, this value may be an overestimate.

Given the difficulty in fitting the GJ 1132 flare data to the \kepler data, we also explored the possibility that the slope of the FFD is always -1, which some data suggest \citep{Loyd18}. We found the that values of the slope and intercept are tightly correlated and that in fact the GJ 1132 values derived from the \tess data are compatible with the \kepler fit. In other words, although the marginal uncertainties for both the slope and intercept appear large (including the slope reaching unphysical values larger than 0), the covariance between them confines their actual values to nearly a straight line that passes through a slope of -1. We conclude that the \tess data are just too sparse to derive a robust fit, but that fit is perfectly consistent with \tess and the hypothesis of a universal FFD slope of -1.



Although we have shown that GJ 1132 b sits firmly on the atmosphere-free side of the \citet{ZahnleCatling17} cosmic shoreline, that should not be taken to mean that the atmosphere has been permanently lost. The cosmic shoreline hypothesis relies heavily on our Solar System, with its unique XUV history and planets that formed from the same protoplanetary disk. Exoplanets may form with significantly more or less volatiles than Venus, Earth, and Mars, and hence may require different amounts of time to lose their atmospheres. Furthermore, the \citet{ZahnleCatling17} formulation is idealized, and alternative theoretical and empirical treatments predict different boundaries between atmospheres and bare rocks \citep[e.g.][]{BertaThompson25,Ih25,Ji25,ChatterjeePierrehumbert26}. We used the original model to be consistent with most previous studies, but the analysis presented above could be used to compare GJ 1132 b to other cosmic shorelines, which could ultimately identify this fundamental boundary in parameter space.

Our analysis of this star should be expanded to additional \jwst targets and provide uncertainties regarding a planet's proximity to the cosmic shoreline. Such a catalog would help interpret data in the context of atmospheric retention, informing JWST observations and help to tune the cosmic shoreline position. A catalog of XUV luminosities and the resultant planet fluxes would also serve as valuable inputs to probabilistic planetary volatile models \citep[\eg][]{KrissansenTottonFortney22,Gialluca24} that can self-consistently track volatile loss.

While light elements such as H, He, and even O can be lost thermally \citep{Zahnle90}, the removal of carbon dioxide is very difficult via thermal processes \citep{KrissansenTottonFortney22}. CO$_2$ could still be lost mechanically through ablation by stellar energetic particles from coronal mass ejections (CMEs), which can strike a planet at relativistic speeds and transfer sufficient momentum to heavy molecules for them reach escape velocity. With the recent claim of the first CME candidate beyond the Solar System \citep{Callingham25}, the community may be on the precipice of directly constraining the CME frequency distribution of main sequence stars. Once the stellar XUV and CME histories can be constrained, the rate of atmospheric loss of volatiles can be properly calculated.

The stellar energetic particles associated with CMEs drive non-thermal escape, which can be influenced by a magnetic field \citep{DriscollBercovici13,Gunell18}, but in principle the loss rate can still be constrained in this case. Since some planets will likely be magnetized, a full accounting of atmospheric retention must include the geodynamo, which is driven by convection in the outer envelopes of cores and is thus set by the temperature difference between the core and mantle \citep{olson2006,DriscollBarnes15}. Thus, a geodynamo prediction must also include the mantle's thermal evolution, which is strongly influenced by the volatile evolution, both in terms of the radiative greenhouse warming of the atmosphere as well as the lubricating effect of water in the mantle that reduces the temperature needed for it to convect \citep{FoleyDriscoll16,McGovernSchubert89,Garcia26}. Thus, a realistic model of permanent atmospheric loss requires a core/mantle/crust/atmosphere/stellar model that couples all these pieces together. Future work should strive toward this goal because predictions for atmospheric loss clearly depend on more sophisticated stellar and planetary modeling than presented here.

\section{Conclusions\label{sec:conclusions}}

We have analyzed the XUV luminosity evolution of GJ 1132 and found that planet b has likely been hit by 50-700 times more XUV photons than Earth, putting its atmosphere at risk of total loss. We compared 2 quiescent XUV models and 1 flaring XUV model and find that all assumptions generate a cumulative XUV flux that the cosmic shoreline hypothesis predicts would result in an atmosphere-free world. We find that the \citet{Ribas05} and \citet{Engle24} models are consistent, but that the latter's uncertainties are about 3 times smaller, suggesting it is the preferred model for stars similar to GJ 1132.

Our framework can be applied to additional planets to help identify those planets that are likely to be devoid of an atmosphere or not. Crucially, including uncertainties in the cumulative XUV flux may enable the validation or falsification of the cosmic shoreline hypothesis with \jwst data. The research described herein motivates a study of the uncertainties in the cumulative XUV flux on planets in the solar neighborhood that may be \jwst targets. Given the M dwarf focus and precision of the \citet{EngleGuinan23} and \citet{Engle24} models, we recommend future work use them to determine the uncertainties in a planet's cumulative flux and proximity to the cosmic shoreline. Moreover, the range of XUV fluxes a planet has received should also be used to interpret observations of these planets' atmospheres, should they possess them.

This research contributes to our understanding of how planets permanently lose their atmospheres, but does not yet produce a realistic prediction for the pathway to total volatile loss. To fully address the processes that lead to permanent atmospheric loss, stellar XUV models like those considered here must be stitched together with planetary models to create a new planetary system model that tracks planetary volatiles between mantle, crust, atmosphere, and space reservoirs, including the role of forced tidal heating in the planetary mantle. Ultimately, the synergy of theory and observation will mature to enable a full understanding of how terrestrial planets totally devolatilize.

\vspace{1cm}

\noindent{\bf Acknowledgments}\\
RB and JB acknowledge support from NASA grant No. 80NSSC23K0261. LA and ES acknowledge support from the NASA Virtual Planetary Laboratory Team through grant number 80NSSC18K0829 and the CHAMPs (Consortium on Habitability and Atmospheres of M-dwarf Planets) team, supported by the National Aeronautics and Space Administration (NASA) under grant Nos. 80NSSC21K0905 and 80NSSC23K1399, issued through the Interdisciplinary Consortia for Astrobiology Research (ICAR) program. This research used Anthropic's Claude Code with the Opus 4.5--4.7 models for generation of some analysis software.

\vspace{1cm}

\noindent{\it Software:} \texttt{alabi} \citep{BirkyBarnes26}, \texttt{astropy} \citep{astropy22}, \texttt{bigplanet} \citep{Barnes20}, \texttt{Claude Code}, \texttt{dynesty} \citep{Speagle20}, \texttt{emcee} \citep{ForemanMackey13}, \texttt{FFD} \citep{Davenport19}, \texttt{lightkurve} \citep{Lightkurve2018}, \texttt{matplotlib} \citep{Hunter07}, \texttt{MaxLEV}, \texttt{multiplanet} \citep{Barnes20}, \texttt{numpy} \citep{harris20}, \texttt{pymultinest} \citep{Buchner14}, \texttt{scipy} \citep{Virtanen20}, \texttt{ultranest} \citep{Buchner21}, \texttt{vaibify}, \texttt{vconverge} \citep{Gialluca24}, \texttt{vplanet} \citep{Barnes20}, \texttt{vplanet\_inference} \citep{Birky25}, \texttt{vspace} \citep{Barnes20}.


\bibliography{gj1132,here}

\appendix

\section{ALABI Kernel and Scaling}

Bayesian inference requires evaluating the likelihood function at
thousands to millions of points in parameter space. When each
likelihood evaluation demands a forward-model simulation that takes
seconds to minutes, direct sampling methods such as MCMC become prohibitively expensive. To overcome this
computational bottleneck, \texttt{alabi} constructs a ``surrogate
model'', a fast, continuously differentiable approximation to the
log-likelihood surface that can be evaluated in microseconds once
trained \citep[see \eg][]{Fleming20}.

In this case, the surrogate model is a Gaussian process (GP), a non-parametric
regression method that defines a probability distribution over
functions, rather than a fit to a function
\citep{Rasmussen06}. Given a set of training points (pairs of
input parameters and their corresponding log-likelihood values), the
GP predicts the log-likelihood at any untested point together with an
uncertainty estimate. This uncertainty quantification identifies where the
surrogate model is poorly constrained, which  guides the selection of
new points during the active learning phase.

A GP is fully specified by two components: a mean function and a
covariance function, the latter commonly called a kernel. The
mean function captures the global trend of the target function (often
set to a constant), while the kernel controls how the GP generalizes
from observed points to unobserved ones. Mathematically, the kernel is a function
$k(\mathbf{x}, \mathbf{x}')$ that returns the covariance between the
GP's predictions at any two points $\mathbf{x}$ and $\mathbf{x}'$ in
parameter space. When two points are close together (relative to the
length scale $\ell$), the kernel returns a value near the output
variance $\sigma^{2}$, meaning the GP expects their log-likelihood
values to be strongly correlated. As the points move farther apart,
the kernel value decays toward zero and the predictions become
independent.

The functional form of the kernel
therefore encodes prior assumptions about the smoothness and
correlation structure of the log-likelihood surface. Choosing a kernel
that is too smooth can wash out genuine features, while a kernel that
is too rough can overfit noise in the training data. Each kernel has a
small number of free parameters (output variance $\sigma^{2}$ and
one or more length scales $\ell$) that must be tuned to the training data. The standard approach is to
maximize the marginal likelihood (sometimes called the model
evidence), which is the probability of the observed training outputs
given the inputs after integrating over all possible functions
consistent with the GP prior. The marginal likelihood naturally
balances model fit against model complexity: a kernel configuration
that fits the training data well but requires an implausibly flexible
function space is penalized relative to a simpler configuration that
fits nearly as well \citep{Rasmussen06}. In practice,
\texttt{alabi} maximizes the log of the marginal likelihood using the
L-BFGS-B gradient-based optimizer \citep{Byrd95}, starting from
multiple initial guesses to reduce the chance of converging to a local
maximum.

In addition to the kernel, the GP's performance depends on how the input
parameters and output log-likelihood values are numerically
represented. For example, the input parameters could span a huge range if masses are expressed in kg, while orbital periods are in days, resulting in log-likelihood values
that are spread out over many orders of magnitude. Without preprocessing, parameters
with large numerical ranges can dominate the distance calculations in
the kernel, effectively rendering the GP insensitive to parameters
with small ranges. Similarly, extreme dynamic range in the output can
cause the GP to focus its predictive power on the largest-magnitude
training points at the expense of accuracy near the likelihood peak,
which is precisely the region that matters most for inference. To mitigate these problems, \alabi can scale the input and output data to place all quantities in a similar absolute range.

Because the optimal kernel and scaling strategy depend on the specific
shape of the log-likelihood surface, which is not known \emph{a
priori}, we leverage \alabi's native functionality to perform a grid search over three kernels and
four scaling strategies. The goal is to
identify the kernel--scaler combination that most accurately
reproduces the true log-likelihood surface, as measured by the
prediction error on a held-out test set. The best configuration is
then used for the full \alabi run.

We considered three stationary kernels: The squared exponential (SE)
kernel,
\begin{equation}
  k_{\mathrm{SE}}(\mathbf{x}, \mathbf{x}') = \sigma^{2}
  \exp\!\left(-\frac{|\mathbf{x} - \mathbf{x}'|^{2}}{2\ell^{2}}\right),
  \label{eq:se_kernel}
\end{equation}
that assumes infinitely differentiable target functions and produces the
smoothest interpolants;  the Mat\'{e}rn 3/2 kernel,
\begin{equation}
  k_{3/2}(\mathbf{x}, \mathbf{x}') = \sigma^{2}
  \left(1 + \frac{\sqrt{3}\,|\mathbf{x} - \mathbf{x}'|}{\ell}\right)
  \exp\!\left(-\frac{\sqrt{3}\,|\mathbf{x} - \mathbf{x}'|}{\ell}\right),
  \label{eq:matern32_kernel}
\end{equation}
that produces once-differentiable sample paths and accommodates rougher
target functions; and the Mat\'{e}rn 5/2 kernel,
\begin{equation}
  k_{5/2}(\mathbf{x}, \mathbf{x}') = \sigma^{2}
  \left(1 + \frac{\sqrt{5}\,|\mathbf{x} - \mathbf{x}'|}{\ell}
  + \frac{5\,|\mathbf{x} - \mathbf{x}'|^{2}}{3\ell^{2}}\right)
  \exp\!\left(-\frac{\sqrt{5}\,|\mathbf{x} - \mathbf{x}'|}{\ell}\right),
  \label{eq:matern52_kernel}
\end{equation}
that is twice-differentiable and occupies the middle ground between the SE
and Mat\'{e}rn 3/2 kernels. In each case, $\sigma^{2}$ is the output
variance and $\ell$ is the characteristic length scale.

In addition to the kernel, the performance of a GP depends on how
the input (model parameters) and output (log-likelihood) values are scaled
before training. We consider three input scalers: no transformation,
min--max scaling to $[0,1]$, and standardization to zero mean and unit
variance. For the output, we add a fourth option: the negative
logarithm, $y \to -\!\ln(-y)$, which compresses the dynamic range of
strongly negative log-likelihoods. The full grid therefore comprises
$3 \times 3 \times 4 = 36$ kernel+scaler combinations.

For each combination, the GP is trained on $N_{\mathrm{train}} = 500$
points drawn from the five-dimensional parameter space via Latin
hypercube sampling (LHS) and evaluated on a separate set of
$N_{\mathrm{test}} = 500$ LHS points. The quality of each
configuration is quantified by the mean squared error (MSE) between
the GP predictions and the true log-likelihood values on the test set,
\begin{equation}
  \mathrm{MSE} = \frac{1}{N_{\mathrm{test}}}
  \sum_{i=1}^{N_{\mathrm{test}}}
  \bigl[\hat{y}(\mathbf{x}_{i}) - y(\mathbf{x}_{i})\bigr]^{2},
  \label{eq:mse}
\end{equation}
where $\hat{y}(\mathbf{x}_{i})$ is the GP prediction and
$y(\mathbf{x}_{i})$ is the forward-model log-likelihood at test point
$\mathbf{x}_{i}$. Both quantities are evaluated in the original
(unscaled) output space to ensure a fair comparison across scaling
strategies.

Table~\ref{tab:grid_search} lists the test MSE for all 36
configurations. Three configurations result in $\mathrm{MSE} = 0.37$, but to more significant figures the best combination is the squared exponential kernel
with standard input scaling and min/max output scaling. During a run, \alabi automatically picks the combination with the lowest MSE, even if other choices are comparable. We did not generate the surrogate model with any of the other combinations.

\begin{deluxetable}{llll}
\tablecaption{GP Kernel and Scaler Grid Search Results\label{tab:grid_search}}
\tablehead{
\colhead{Kernel} & \colhead{Input Scaler} & \colhead{Output Scaler} & \colhead{Test MSE}
}
\startdata
Squared Exponential & None          & None           &  11.44 \\
Squared Exponential & None          & Negative Log   &  90.29 \\
Squared Exponential & None          & Min-Max        &   0.37 \\
Squared Exponential & None          & Standard       &   2.90 \\
Squared Exponential & Min-Max       & None           &  10.23 \\
Squared Exponential & Min-Max       & Negative Log   &  57.28 \\
Squared Exponential & Min-Max       & Min-Max        &   0.37 \\
Squared Exponential & Min-Max       & Standard       &   3.18 \\
Squared Exponential & Standard      & None           &   4.10 \\
Squared Exponential & Standard      & Negative Log   &  58.22 \\
Squared Exponential & Standard      & Min-Max        &   0.37 \\
Squared Exponential & Standard      & Standard       &   3.18 \\
Mat\'{e}rn 3/2      & None          & None           &   2.31 \\
Mat\'{e}rn 3/2      & None          & Negative Log   &  11.52 \\
Mat\'{e}rn 3/2      & None          & Min-Max        &   1.07 \\
Mat\'{e}rn 3/2      & None          & Standard       &   1.00 \\
Mat\'{e}rn 3/2      & Min-Max       & None           &   2.57 \\
Mat\'{e}rn 3/2      & Min-Max       & Negative Log   &  11.53 \\
Mat\'{e}rn 3/2      & Min-Max       & Min-Max        &   1.07 \\
Mat\'{e}rn 3/2      & Min-Max       & Standard       &   0.99 \\
Mat\'{e}rn 3/2      & Standard      & None           &   2.58 \\
Mat\'{e}rn 3/2      & Standard      & Negative Log   &  11.52 \\
Mat\'{e}rn 3/2      & Standard      & Min-Max        &   1.07 \\
Mat\'{e}rn 3/2      & Standard      & Standard       &   0.99 \\
Mat\'{e}rn 5/2      & None          & None           &   1.92 \\
Mat\'{e}rn 5/2      & None          & Negative Log   &  60.04 \\
Mat\'{e}rn 5/2      & None          & Min-Max        &   0.51 \\
Mat\'{e}rn 5/2      & None          & Standard       &   0.47 \\
Mat\'{e}rn 5/2      & Min-Max       & None           &   2.16 \\
Mat\'{e}rn 5/2      & Min-Max       & Negative Log   &  15.26 \\
Mat\'{e}rn 5/2      & Min-Max       & Min-Max        &   0.51 \\
Mat\'{e}rn 5/2      & Min-Max       & Standard       &   0.47 \\
Mat\'{e}rn 5/2      & Standard      & None           &   2.16 \\
Mat\'{e}rn 5/2      & Standard      & Negative Log   &  15.24 \\
Mat\'{e}rn 5/2      & Standard      & Min-Max        &   0.51 \\
Mat\'{e}rn 5/2      & Standard      & Standard       &   0.47 \\
\enddata
\end{deluxetable}

\section{ALABI Training and Active Learning}

With the kernel and scaling methods selected, we must also determine the best approach to generating the GP surrogate model. The key variations include the number of training points $N_{init}$, the number of active learning iterations, setting and fitting white noise, and the frequency of GP hyperparameter optimizations $f_{opt}$. We considered 5 different tests with 500, 1500, and 3000 training points, GP optimization every 10 or 50 iterations, and fitting or not fitting white noise.

Since the actual posteriors are not known {\it a priori} and the kernel+scalings might not work in the final run, we must also carefully examine the \alabi output to confirm that the algorithm did not show spurious behavior and that the surrogate model does produce a realistic posterior. We assess the viability of each permutation by first computing the normalized root mean squared error,
\begin{equation}
\label{eq:nrmse}
\mathrm{NRMSE} =  100\sqrt{{\rm MSE}_{\rm test}/{\rm Var}(y_{\rm train})},
\end{equation}
where MSE$_{\rm test}$ is the mean squared error, and Var($y_{\rm train}$) is the variance of the training data with 500 held-out test points (100x number of dimensions), as recommended by \citet{BirkyBarnes26}. We compute the NRMSE  throughout active learning, with lower values indicating better convergence. We also make a more qualitative assessment of the stability of the surrogate model to active learning iterations by comparing the ratio of the testing and training data NRMSE's with iteration number and by checking for an increasing NRMSE. Ratios that remain nearly constant and below 2 and for which the test NRMSE does not significantly increase are deemed stable, \ie they did not overfit.

Table \ref{tab:convergence} shows the results of 5 different trials to assess accuracy and stability, using the exponential squared kernel with standard input scaling and min/max output scaling. We found that 500 training points (100x the number of dimensions) was insufficient to successfully converge the posterior -- the training data was too sparse and the length scale hyperparameter became too large to properly resolve the likelihood peak. At 1500 training points, however, the NRMSE drops to nearly 1\% and appears stable during active learning. Further testing showed that hyperparameter optimization every 10 iterations tended to result in either unstable active learning or a larger NRMSE. We also found that increasing the training data set to 3000 points resulted in only modest improvements in the NRMSE, but at a factor of 3 increase in computational time due to both the generation of more forward models, but also because GP optimization scales with the total number of points.

\begin{table}
\centering
\caption{GP surrogate convergence across training configurations.}
\label{tab:convergence}
\begin{tabular}{ccccccc}
\hline
Run & $N_{\rm init}$ & $f_{\rm opt}$ & White Noise & Best NRMSE (\%) & Final NRMSE (\%) & Stable \\
\hline
1 & 500  & 50 & No  &  9.00 & 96.64 & No  \\
2 & 1500 & 50 & No  &  2.63 &  2.73 & Yes \\
3 & 1500 & 10 & Yes &  4.21 &  4.27 & Yes \\
4 & 3000 & 10 & No  &  2.38 & 13.24 & No  \\
5 & 3000 & 50 & No  &  2.38 &  2.41 & Yes \\
\hline
\end{tabular}
\end{table}

\begin{figure}
    \includegraphics[width=\textwidth]{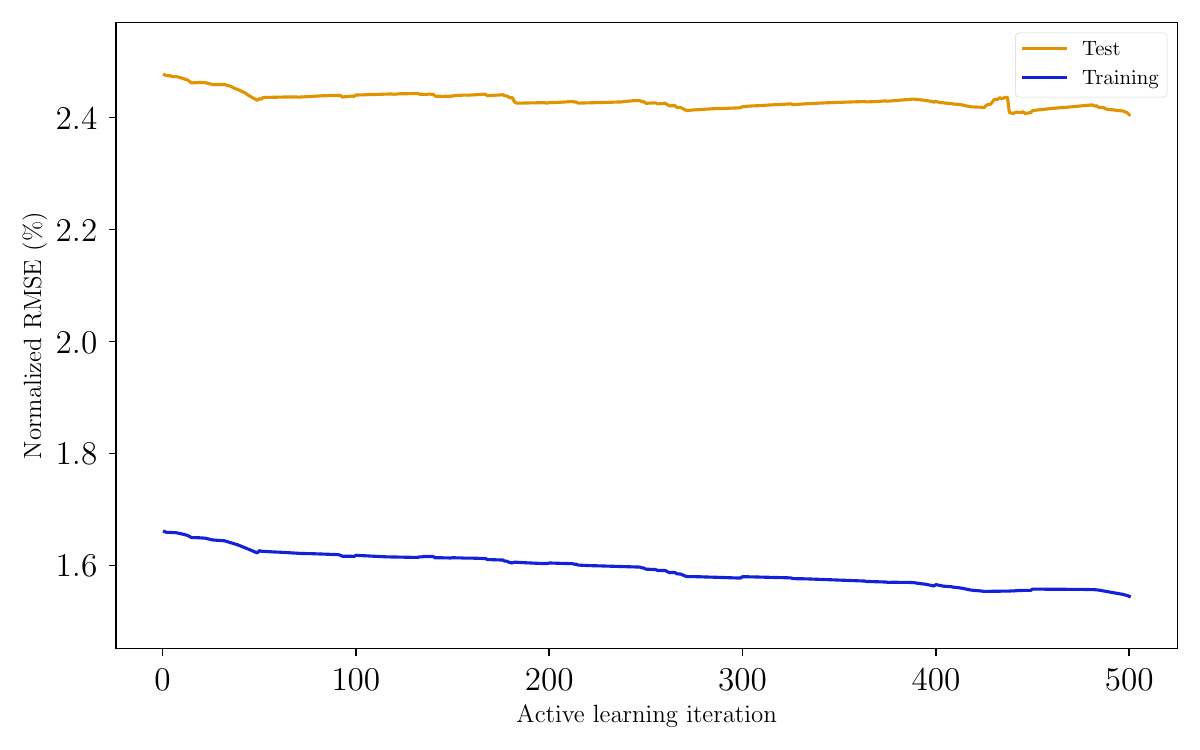}
    \caption{Normalized root mean squared error, see Eq.~(\ref{eq:nrmse}), during active learning of Run \#5 in Table \ref{tab:convergence}. The orange curve is the NRMSE during active learning, while the blue curve is the NRMSE  for the test data with 500 simulations randomly held out each iteration.}
    \label{fig:nrmse}
\end{figure}

We thus conclude that Run \#2 is the optimal configuration for this problem because its NRMSE is nearly the same as Run \#5, but it required 3x less computational time to run. However, because we did run the $N_{init} = 3000$, $f_{opt} = 50$ case, we presented the posterior for this surrogate model in Fig.~\ref{fig:XUVPosteriors}. In Fig.~\ref{fig:nrmse} we show the evolution of the NRMSE during active learning for this problem. Note that although the training data are fixed during active learning, its NRMSE can vary due to the randomly held out simulations. In this case, the training NRMSE experiences only a slight decline over the 500 active learning iterations.

\end{document}